\documentclass[aps,prl,floatfix,superscriptaddress,preprint]{revtex4-1} 
\usepackage{graphicx}
\usepackage{amsmath,amsfonts,amssymb}
\usepackage{color}
\usepackage{multibib}

\usepackage{pdfpages}

\makeatletter
\AtBeginDocument{\let\LS@rot\@undefined}
\makeatother

\begin{document}
\title{Breakdown of Magnons in a Strongly Spin-Orbital Coupled Magnet}
\date{\today}
\begin{abstract}
The description of quantized collective excitations stands as a landmark in the quantum theory of condensed matter. A prominent example occurs in conventional magnets, which support bosonic magnons - quantized harmonic fluctuations of the ordered spins. In striking contrast is the recent discovery that strongly spin-orbital coupled magnets, such as $\alpha$-RuCl$_3$, may display a broad excitation continuum inconsistent with conventional magnons.
Due to incomplete knowledge of the underlying interactions unraveling the nature of this continuum remains challenging. 
The most discussed explanation refers to a coherent continuum of fractional excitations analogous to the celebrated Kitaev spin liquid.
Here we present a more general scenario.
We propose that the observed continuum represents incoherent excitations originating from strong magnetic anharmoniticity that naturally occurs in such materials. This scenario fully explains the observed inelastic magnetic response of $\alpha$-RuCl$_3$ and reveals the presence of nontrivial excitations in such materials extending well beyond the Kitaev state. \end{abstract}

\author{Stephen M. Winter}
\email{winter@physik.uni-frankfurt.de}
\affiliation{Institut f\"ur Theoretische Physik, Goethe-Universit\"at Frankfurt,
Max-von-Laue-Strasse 1, 60438 Frankfurt am Main, Germany}
\author{Kira Riedl}
\affiliation{Institut f\"ur Theoretische Physik, Goethe-Universit\"at Frankfurt,
Max-von-Laue-Strasse 1, 60438 Frankfurt am Main, Germany}
\author{Pavel A. Maksimov}
\affiliation{Department of Physics and Astronomy, University of California, Irvine, California 92697, USA}
\author{Alexander L. Chernyshev}
\affiliation{Department of Physics and Astronomy, University of California, Irvine, California 92697, USA}
\author{Andreas Honecker}
\affiliation{Laboratoire de Physique Th\'eorique et Mod\'elisation, CNRS UMR 8089,
Universit\'e de Cergy-Pontoise, 95302 Cergy-Pontoise Cedex, France}
\author{Roser Valent{\'\i}}
\affiliation{Institut f\"ur Theoretische Physik, Goethe-Universit\"at Frankfurt,
Max-von-Laue-Strasse 1, 60438 Frankfurt am Main, Germany}

\maketitle

From magnons in ordered magnets to  phonons in periodic crystals, the appearance of bosonic collective excitations is ubiquitous in condensed phases of matter \cite{lifshitz2013statistical}. For this reason, special attention is given to those states that support more exotic collective modes, for which the conventional paradigm breaks down. In the context of magnetic phases, the breakdown of magnons is commonly thought to require closeness to an unconventional state such as a quantum spin liquid \cite{balents2010spin,PhysRevLett.86.1335,han2012fractionalized}. A notable example occurs in Kitaev's exactly solvable honeycomb model \cite{kitaev2006anyons}, for which strongly anisotropic and bond-dependent interactions fractionalize conventional spin excitations into Majorana spinons and fluxes. This Kitaev state has recently risen to prominence due to the suggestion that it may be realized in heavy metal $4d^5$ and $5d^5$ insulators via a specific interplay between the crystal field and strong spin-orbit coupling \cite{PhysRevLett.102.017205} and, consequently, a variety of candidate materials based on Ir$^{4+}$ and Ru$^{3+}$ have been intensively explored \cite{rau2016spin}. Encouragingly,  evidence of a continuum of magnetic excitations that is inconsistent with conventional magnons was found in the majority of such materials,
including the two-dimensional (2D) honeycomb Na$_2$IrO$_3$ \cite{gretarsson2013magnetic,chun2015direct} and $\alpha$-RuCl$_3$ \cite{sandilands2015scattering,nasu2016fermionic,banerjee2016neutron,banerjee2016proximate,incarnation}, as well as the
3D analogues $\beta$-,$\gamma$-Li$_2$IrO$_3$ \cite{glamazda2016raman}, despite all of them having magnetically ordered ground states.

While the observed excitation continua in these systems have been interpreted in terms of signatures of the Kitaev state, the low-symmetry crystalline environment of the real materials also allows various additional interactions beyond Kitaev's model \cite{chaloupka2010kitaev,rau2014generic,PhysRevB.93.214431}, which are thought to be large based on both experimental \cite{sears2015magnetic,PhysRevB.94.064435} and theoretical \cite{PhysRevB.91.245134,0295-5075-112-6-67004,PhysRevB.93.214431}
considerations. In this sense, understanding the mechanism for the breakdown of magnons and the appearance of a broad
continuum of magnetic excitations remains a key challenge.

In this work, we study a representative case $\alpha$-RuCl$_3$, which forms a layered 2D honeycomb lattice and displays zigzag magnetic order below $T_\text{N} \sim 7$ K \cite{banerjee2016proximate,PhysRevB.93.134423,banerjee2016neutron}. We specifically address the recent inelastic
neutron scattering (INS) measurements, which have revealed low-energy magnons \cite{ranneutron} coexisting with an intense excitation continuum \cite{banerjee2016neutron}. The latter continuum possesses a distinctive six-fold star shape in momentum space, and large intensity at the 2D $\Gamma$-point over a wide energy range $E = 2 - 15$ meV \cite{banerjee2016neutron}. 
To resolve the nature of this continuum, we take two complementary approaches. We first theoretically investigate the neutron spectra over a range of relevant magnetic interactions in order to 
determine the correct spin Hamiltonian for $\alpha$-RuCl$_3$, which has been a subject
of intense recent discussion \cite{kim2016crystal,PhysRevB.93.214431,yadav2016spin,wanginteractions,houinteractions}.
Second, we identify the conditions that lead to the
breakdown of  conventional magnons in the presence of strongly anisotropic and frustrated interactions, revealing that nontrivial
excitations naturally persist well beyond the Kitaev spin liquid.
 \\
\\
\textbf{Results} \\
\textbf{The model.}
Based on previous {\it ab initio} studies \cite{kim2016crystal,PhysRevB.93.214431,yadav2016spin,wanginteractions,houinteractions},
the largest terms in the spin Hamiltonian of $\alpha$-RuCl$_3$ are generally expected to include
nearest neighbour Heisenberg $J_1$, Kitaev $K_1$, and off-diagonal $\Gamma_1$
couplings, supplemented by a 3rd neighbour Heisenberg $J_3$ term: \begin{align}
\mathcal{H} =& \  \sum_{\langle i,j\rangle} J_1 \ \mathbf{S}_i \cdot \mathbf{S}_j + K_1 S_i^\gamma S_j^\gamma + \Gamma_1 (S_i^\alpha S_j^\beta + S_i^\beta S_j^\alpha) \nonumber \\  & \ +  \sum_{\langle\langle\langle i,j\rangle\rangle\rangle} J_3\  \mathbf{S}_i \cdot \mathbf{S}_j \label{eqn-1}
\end{align}
where $\langle i,j\rangle$ and $\langle\langle\langle i,j\rangle\rangle\rangle$ refer to summation over first and third neighbour bonds, respectively (see Fig.~1). The bond-dependent variables $\{\alpha,\beta,\gamma\}$ distinguish the three types of first neighbour bonds, with $\{\alpha,\beta,\gamma\} = \{y,z,x\}, \{z,x,y\}$, and $\{x,y,z\}$ for the X-, Y-, and Z-bonds, respectively. The third neighbour interactions are bond-independent. The phase diagram of this model has been discussed previously \cite{rau2014generic,katukuri2014kitaev,PhysRevB.93.214431,yadav2016spin}, and
is further detailed in Supplementary Note 1; here we review the key aspects.

In the
limit $J_1= \Gamma_1=J_3=0$, the ground state is a gapless
Z$_2$ spin liquid for either positive or negative $K_1$, as demonstrated in Kitaev's seminal work \cite{kitaev2006anyons}. Small perturbations from
the pure $K_1$ limit may induce various magnetically ordered states, such as the zigzag antiferromagnetic state observed in $\alpha$-RuCl$_3$ and  shown in Fig.~1. The simplest perturbation is the introduction of a finite $J_1$, which yields the well-studied ($J_1,K_1$) nearest
neighbour Heisenberg-Kitaev (nnHK) model. This model hosts zigzag order in the region $K_1 >
0, J_1 <0$, as discussed in Supplementary Note 1. Accordingly, previous analysis of the powder INS experiments within the context of the nnHK model \cite{banerjee2016proximate}, suggested that $K_1
\sim +7$ meV, and $|J_1/K_1| \sim 0.3-0.7$ for $\alpha$-RuCl$_3$. On this basis, the excitation continua observed experimentally were initially interpreted in terms of proximity to
the antiferromagnetic (AFM) $K_1 > 0$ spin liquid \cite{banerjee2016proximate,banerjee2016neutron}.  However, the further consideration of finite 
$\Gamma_1$ and $J_3$ interactions in Eq.~\eqref{eqn-1} significantly expands the
experimentally relevant region, as both interactions generally stabilize zigzag
order. Indeed, recent {\it ab initio} studies \cite{yadav2016spin,kim2016crystal,PhysRevB.93.214431,wanginteractions,houinteractions} have
suggested that the zigzag order in $\alpha$-RuCl$_3$ emerges from $J_1 \sim 0, K_1 < 0, \Gamma_1 > 0$, and $J_3 >0$, with
$|\Gamma_1/K_1| \sim 0.5 - 1.0$, and $|J_3/K_1| \sim 0.1-0.5$, as reviewed in Supplementary Note 2. That is, $K_1$ is ferromagnetic, and supplemented by significant $\Gamma_1$ and $J_3$ interactions. Such interactions would represent large deviations from both Kitaev's original model and the region
identified by initial experimental analysis. Before discussing the origin of the excitation continua, it is therefore crucial to first pinpoint the relevant interaction parameters.

In order to address this issue directly, we have computed the neutron
scattering intensity $\mathcal{I}(\mathbf{k},\omega)$ for a variety of
interactions within the zigzag ordered phase via both linear spin-wave theory
(LSWT) and exact diagonalization (ED). For the latter case, we combine results
from various periodic 20- and 24-site clusters compatible with the zigzag state
in order to probe a wider range of $\mathbf{k}$-points (see Methods section).
Full results for the complete range of models are presented in Supplementary Note 5. Here, we highlight the key results for two representative sets of interactions. Within the ($J_1,K_1$) nnHK model, we focus on \mbox{Model 1} ($J_1 = -2.2,K_1=+7.4$ meV; $ |J_1/K_1|=0.3$), which lies on the
border of the region initially identified in Ref.~\cite{banerjee2016neutron}, close to the spin liquid. 
Beyond the nnHK model, we consider
\mbox{Model 2}  ($J_1 = -0.5,K_1=-5.0,\Gamma_1=+2.5,J_3=+0.5$ meV) for which
parameters have been  guided by recent {\it ab initio} studies \cite{yadav2016spin,kim2016crystal,PhysRevB.93.214431,wanginteractions,houinteractions}, and further optimized to improve agreement with the experimental spectra. Results for Models 1 and 2 are first presented in Fig.~\ref{fig-spectralAFM} and Fig.~\ref{fig-spectralFM}, which show detailed $\omega$- and $\mathbf{k}$-dependence of $\mathcal{I}(\mathbf{k},\omega)$, along with the evolution of the spectra upon changing parameters towards the $K_1 >0$ or $K_1 < 0$ spin liquid regions.

\textbf{Nearest neighbour Heisenberg-Kitaev (nnHK) model.}
We begin by analyzing the spectra  $\mathcal{I}(\mathbf{k},\omega)$ within the zigzag phase of the $(J_1,K_1)$ nnHK model, starting with Model 1
(Fig.~\ref{fig-spectralAFM}).
 Despite proximity to the spin liquid, the ED
calculations on Model 1 (Fig.~\ref{fig-spectralAFM}b) show sharp dispersive modes appearing over the majority of the
Brillouin zone that are consistent with the conventional magnons of
LSWT (Fig.~\ref{fig-spectralAFM}a). Indeed, for energies below the main spin-wave branch ($\omega = 1.3-2.3$ meV), intensity is localized around the M- and Y-points, corresponding to the pseudo-Goldstone modes of the zigzag order (Fig.~\ref{fig-spectralAFM}c). ED calculations show clear spin-wave cones emerging from such points and extending to higher energies. Large deviations from LSWT are observed only for the highest energy excitations, which appear near the 2D $\Gamma$-point for energies $\omega > 5$ meV. Here, the ED calculations display a broad continuum-like feature centred at $\omega \sim K_1$ that resembles the response expected for the $K_1 > 0$ Kitaev spin liquid, as shown in Fig.~\ref{fig-spectralAFM}d. However, comparison with the experimental $\mathcal{I}(\Gamma,\omega)$ shows poor agreement; while the experimental intensity extends from $2-15$ meV, the ED results for Model 1 predict intensity only at high energies $> 5$ meV. Indeed, the evolution of the $\Gamma$-point intensity with
$|J_1/K_1|$ is shown in Fig.~\ref{fig-spectralAFM}e. On approaching the $K_1 >0$
spin liquid by decreasing $|J_1/K_1|$, excitations at the $\Gamma$-point shift to higher
energy, such that none of the parameters in the vicinity of the spin liquid
reproduce the experimental $\omega$-dependence of $\mathcal{I}(\Gamma,\omega)$. Similar conclusions can also be drawn from recent DMRG
studies of the nnHK model \cite{frank2017}. We therefore conclude that 
the broad features observed experimentally in $\mathcal{I}(\Gamma,\omega)$ at relatively low energies \cite{banerjee2016neutron} are incompatible with the nnHK model with $J_1 < 0$ and $K_1>0$.

\textbf{Extended {\it \textbf{ab initio}} guided model.}
In order to treat the effect of interactions beyond the nnHK model, we consider
now the {\it ab initio} guided Model 2. In
contrast to Model 1, ED calculations on Model 2 (Fig.~\ref{fig-spectralFM}b), show large deviations from
standard LSWT (Fig.~\ref{fig-spectralFM}a) over a wide range of $\mathbf{k}$ and $\omega$. This model reproduces many of the experimental spectral features \cite{ranneutron,banerjee2016neutron}.
 In particular, sharp single-magnon-like peaks appear only near the
pseudo-Goldstone modes at the M- and Y-points. Elsewhere in the Brillouin zone, broad continuum-like
features are observed within the ED resolution. As demonstrated in Fig.~\ref{fig-spectralFM}c, we find significant intensity at low energies ($\omega < 2.3$ meV), at both the $\Gamma$- and
(M,Y)-points. For the intermediate energy region ($\omega = 5.5 - 8.5$ meV), $\mathcal{I}(\mathbf{k})$
resembles the six-fold star shape observed in Ref.~\cite{banerjee2016neutron}. At higher
energies ($\omega > 10.5$ meV) scattering intensity is mainly located at the
$\Gamma$-point, also in accord with experiment. Furthermore, the ED results for the $\Gamma$-point intensity $\mathcal{I}(\Gamma, \omega)$ show a broad range of
excitations peaked around 4 and 6 meV, and extending up to $\sim 15$ meV (Fig.~\ref{fig-spectralFM}d). Therefore, ED
calculations on Model 2 reproduce all of the main experimental spectral 
features, validating the range of interactions indicated by {\it ab initio}
calculations. The only aspect that is not quantitatively reproduced within the Model 2 is the magnitude of the gap at the M-point ($\sim 0.8$ meV at the level of LSWT vs. $\sim$ 2 meV experimentally \cite{ranneutron,banerjee2016proximate}). This discrepancy may result from deviations from $C_3$ symmetry, which are allowed within the $C2/m$ space group \cite{PhysRevB.93.214431,PhysRevB.92.235119}, but not considered here for simplicity (see Supplementary Figure 11). Interestingly, the spectral features at the $\Gamma$-point become
dramatically sharper on approaching the $K_1<0$ spin liquid, as shown in the evolution of $\mathcal{I}(\Gamma,\omega)$ with the ratio $|\Gamma_1/K_1|$ (Fig.~\ref{fig-spectralFM}e). This result reveals that the broad continuum may not be directly
associated with a proximity to the Kitaev state.

\textbf{Magnon stability beyond LSWT.}
To gain further insight into the reason for such a drastic contrast between the stability of magnons in Models 1 and 2, it is useful to consider possible magnon decay channels in the zigzag ordered phase. At the level of
LSWT, the spin-wave Hamiltonian is truncated at quadratic order, and can be written $\mathcal{H}_2 =
\sum_\mathbf{k,m} \epsilon_{\mathbf{k},m} \ a_{\mathbf{k},m}^\dagger
a_{\mathbf{k},m}$ in terms of magnon creation (annihilation) operators
$a^\dagger$ ($a$), where $\epsilon_{\mathbf{k},m}$ denotes the dispersion of
the $m$th magnon band. In this harmonic approximation, magnons represent
sharp, well-defined excitations. However, when higher order anharmonic terms are included, the total magnon number $N_{\text{tot}}=\sum_{\mathbf{k},m} a_{\mathbf{k},m}^\dagger a_{\mathbf{k},m}$ is typically not a conserved quantity, such that the stability of magnons is not guaranteed beyond quadratic order. Quantum fluctuations associated with the higher-order anharmonic decay
terms may mix sharp single-magnon modes with the
multi-magnon continuum \cite{harris1971dynamics,chernyshev2006magnon,zhitomirsky2013colloquium}. Similar considerations also apply to the breakdown of other collective modes, such as phonons in anharmonic crystals \cite{PhysRevB.2.1172, kosevich}. From this perspective, a large decay rate is expected for any single magnon mode that is energetically degenerate with the multi-particle continuum, unless there are specific symmetries guaranteeing that the two do not couple. It is therefore useful to consider the prerequisites for magnon breakdown in the presence of the strongly anisotropic interactions of Eq.~\eqref{eqn-1}. 

\textbf{Magnon decay channels for the nnHK model.} We first examine the stability of magnons in the nnHK model. For pure $J_1$ and $K_1$ interactions, the total spin projections $S_{\text{tot}}^\gamma = \sum_i
S_i^\gamma$ are conserved along the cubic axes $\gamma = \{x,y,z\}$ modulo two.
Since the ordered moment also lies along one of the cubic axes in the zigzag
phase \cite{PhysRevB.94.064435,PhysRevB.94.085109}
(see Fig.~\ref{fig-AFMb}c), the
possible magnon decay channels are restricted. In the local picture,
the relevant quantum fluctuations are local singlet $S_i^x S_j^x |\uparrow
\downarrow\rangle = |\downarrow \uparrow\rangle $ and triplet $S_i^x S_j^x
|\uparrow \uparrow\rangle = |\downarrow \downarrow\rangle $ fluctuations shown
in Fig.~\ref{fig-AFMb}a, with $\Delta S_\text{tot}^z = 0$ and 2, respectively. In
the magnon picture, the Hamiltonian can only contain even order terms (i.e.
$\mathcal{H} = \mathcal{H}_2+\mathcal{H}_4+...$), analogous to conventional
Heisenberg antiferromagnets with collinear ordered spins \cite{harris1971dynamics,zhitomirsky2013colloquium}. For example, the fourth order decay process due to $\mathcal{H}_4$ mixes the one-magnon states with the
three-magnon continuum ($\Delta N_{\text{tot}} = \pm 2$), where
\begin{align}
\mathcal{H}_4 =& \sum_{\mathbf{1}-\mathbf{4}} V^{\mathbf{4}}_{\mathbf{123}} \ a^\dagger_{\mathbf{1}} a^\dagger_{\mathbf{2}}a^\dagger_{\mathbf{3}}a_{\mathbf{4}}\ \delta(\mathbf{k}_1+\mathbf{k}_2+\mathbf{k}_3-\mathbf{k}_4)+H.c.
\end{align}
Here, the bold index ($\mathbf{n}\equiv \mathbf{k}_n,m_n$) labels both momentum and magnon band. This process is pictured in Fig.~\ref{fig-AFMb}b. As noted above, the effect of such terms depends crucially on the availability of low-energy three-magnon states in which to decay.

The density of three-magnon states for Model 1 is  shown in
Fig.~\ref{fig-AFMb}d, based on the one-magnon dispersions obtained in LSWT. At each $\mathbf{k}$-point, the lowest energy
three-magnon state $a_{\mathbf{q}_1}^\dagger a_{\mathbf{q}_2}^\dagger
a_{\mathbf{q}_3}^\dagger |0\rangle$, (with
$\mathbf{q}_1+\mathbf{q}_2+\mathbf{q}_3 = \mathbf{k}$) is obtained by placing
two particles in the pseudo-Goldstone modes at opposite M-points ($\mathbf{q}_1
+ \mathbf{q}_2 = 0$), and the third particle at $\mathbf{q}_3 = \mathbf{k}$,
with total energy $E_3^{\text{min}} (\mathbf{k}) = \epsilon_{\mathbf{k},1} +
2\epsilon_{\text{M},1}$. This implies $E_3^{\text{min}} (\mathbf{k})  \geq
\epsilon_{\mathbf{k},1}$. That is, the three-magnon states lie above the lowest
one-magnon band at every $\mathbf{k}$-point. As a result, every magnon in the lowest band remains kinetically stable, due to the absence of low-energy three-particle states in which to decay. Precisely this condition ensures the stability of low-energy magnons in conventional isotropic antiferromagnets, and explains the sharp magnon-like peaks observed in Fig.~\ref{fig-spectralAFM}b for Model 1. 
Strong spectral broadening in the nnHK model can occur only for high-lying
excitations with $\epsilon_{\mathbf{k},m} > \epsilon_{\mathbf{k},1}=E_3^{\text{min}}$,
where the density of three-magnon states is finite, such as at the 2D $\Gamma$-point. On approaching the spin liquid (at $J_1/K_1$ = 0), this condition is relaxed due to the vanishing dispersion of the lowest magnon band (i.e.
$\epsilon_{\mathbf{k},1}\rightarrow 0$), which corresponds to a vanishing energy cost the singlet fluctuations shown on the left of Fig.~\ref{fig-AFMb}a. The relevant fluctuations in the limit $J_1/K_1 \rightarrow 0$ therefore correspond to $\Delta N_\text{tot} = \pm 2$. For other values of $J_1/K_1$, the majority of magnons are expected to remain stable due to the absence of low-energy three-magnon states.

\textbf{Magnon decay channels for the extended model.} In Model 2, the character of the quantum fluctuations away from zigzag order is notably different
(Fig.~\ref{fig-FMb}). The finite $\Gamma_1$ interaction reduces the local symmetry and
leads to rotation of the ordered moments away from the cubic axes \cite{PhysRevB.94.064435,PhysRevB.94.085109} (Fig.~\ref{fig-FMb}c).
 In the local picture, this allows additional
single-spin fluctuations $S_i^xS_j^z|\uparrow\uparrow\rangle = |\downarrow
\uparrow\rangle$ (Fig. \ref{fig-FMb}a), which correspond to odd-order
anharmonic terms $\mathcal{H}_3,\mathcal{H}_5,...$ in the magnon Hamiltonian,
where \cite{chernyshev2006magnon,zhitomirsky2013colloquium}:
\begin{align} \mathcal{H}_3 =& \ \sum_{\mathbf{1-3}}
\Lambda_\mathbf{12}^\mathbf{3} \ a^\dagger_{\mathbf{1}}
a^\dagger_{\mathbf{2}}a_{\mathbf{3}}\
\delta(\mathbf{k}_1+\mathbf{k}_2-\mathbf{k}_3) + H.c. \label{eqn-3}
\end{align}
At lowest order, such terms mix the single-magnon states with the two-magnon continuum ($\Delta N_{\text{tot}} = \pm 1$), via the scattering process depicted in Fig.~\ref{fig-FMb}b. The density of two-magnon states is shown in Fig.~\ref{fig-FMb}d, for the zigzag domain with $\mathbf{Q}$ = Y. In this case, at each $\mathbf{k}$-point the lowest energy two-magnon state $a_{\mathbf{q}_1}^\dagger a_{\mathbf{q}_2}^\dagger |0\rangle$  is obtained by placing one particle in the pseudo-Goldstone mode at an M-point, and the second particle at $\mathbf{q}_2 = \mathbf{k}-$M, with total energy $E_2^{\text{min}}(\mathbf{k}) = \epsilon_{\mathbf{k}-\text{M}} + \epsilon_\text{M} \neq E_3^{\text{min}}$. 
It should be emphasized that this condition differs from that of a conventional Heisenberg antiferromagnet (for which $E_2^{\text{min}} = E_3^{\text{min}}$) \cite{zhitomirsky2013colloquium}. In the case of Model 2, the difference is directly related to the strong anisotropic $K_1$ and $\Gamma_1$ interactions, which shift the pseudo-Goldstone modes to the M-points, such that only high energy magnons remain at the $\Gamma$-point or ordering wavevector $\mathbf{Q}$ \cite{chaloupka2015hidden}. This shift therefore leads to an offset of the low-energy even and odd magnon states in $\mathbf{k}$-space such that
$E_2^{\text{min}}(\mathbf{k}) < \epsilon_{\mathbf{k},1}$ over a wide region of the
Brillouin zone; there are many two-magnon states with equal or lower energy
than the one-magnon states. 
Provided there is a finite $\Gamma_1$, the
spontaneous decay of single magnons into the two-particle continuum is
therefore allowed even for the lowest magnon band. The decay rate is expected to be particularly large near the zone center, which represents a minimum in $E_2^\text{min}$. 
Similar kinematic conditions may also occur in other systems \cite{zhitomirsky2013colloquium,PhysRevB.93.235130}.
For Model 2, the
pseudo-Goldstone magnons near the M-points remain coherent due to the absence
of low-energy two particle states in which to decay (Fig.~\ref{fig-FMb}d). This explains the experimental observation of sharp magnon-like modes near the M-points \cite{ranneutron}. In
contrast, the magnon bands in the remainder of the Brillouin zone directly
overlap with the two-particle continuum. It is therefore natural to anticipate a large decay rate even for the lowest magnon bands. 

To confirm this idea, we have computed the three-magnon interactions and decay rates
for all magnon bands for Model 2 using the self-consistent imaginary Dyson equation (iDE) approach \cite{PhysRevB.94.140407}. Within this approach, it is assumed that the real part of the magnon self-energy is already captured by the LSWT parameters, while the imaginary part is obtained self-consistently (see Methods and Supplementary Note 3). The iDE approach therefore represents an extension of LSWT, in which the one-magnon excitations are broadened according to the momentum and band-dependent decay rate $\gamma_{\mathbf{k},n}$, while other contributions to the neutron intensity from multi-magnon excitations are also absent \cite{PhysRevB.88.094407}. As a result, comparison of LSWT, ED, and iDE results (Fig.~\ref{fig_ide}) allows for the identification of the origin of different contributions to the spectra. 

The predicted neutron scattering intensity within the iDE approach (Fig.~\ref{fig_ide}b) captures many of the most notable features that are observed in the ED and experimental data,
showing a significant improvement over the LSWT results (Fig.~\ref{fig_ide}a).
First, there is an almost complete washout of the two high-energy one-magnon modes due to strong decays. This implies that the higher-energy features $> 4$ meV appearing in ED are primarily multi-magnon in character (including the 6 meV peak at the $\Gamma$-point). The appearance of these higher energy features in the inelastic neutron response may arise partly from direct contributions from the broadened two-magnon continuum via the longitudinal component of the structure factor, which is not included in the iDE approach (see Supplementary Note 3). Second, the broadening of the two lower magnon bands in the iDE results and the resultant variation of their intensities are in a close agreement with the exact diagonalization -- particularly in a wide region near the $\Gamma$ point (see also Supplementary Figure~5). These are precisely the features with which the LSWT results were most incompatible. Over much of the Brillouin zone -- and especially for the higher magnon bands -- the computed $\gamma_{\mathbf{k},n}$ is on the same scale as the one-magnon bandwidth, confirming the absence of coherent magnons.

\textbf{Discussion.}

The general requirements for strong two-magnon decays are less restrictive than a proximity to a spin-liquid state. Indeed, a large decay rate is ensured by the following three conditions: large anisotropic interactions, deviation of the ordered
moments away from the high-symmetry axes, and strong overlap of the one-magnon states with the multi-magnon continuum (see Supplementary Note 3). Of these, the first two conditions ensure that the scattering vertex $\Lambda_\mathbf{12}^\mathbf{3}$ is large --  of the order of the underlying interactions, i.e. $\Lambda_\mathbf{12}^\mathbf{3} \sim \mathcal{O}(K_1,\Gamma_1)$. For $\alpha$-RuCl$_3$, the strong overlap with the multi-magnon continuum is ensured by shifting of the low-energy magnons away from the $\Gamma$-point. Since the bottom of the two-magnon continuum must always have an energetic minimum at the $\Gamma$-point, the shifting of the pseudo-Goldstone modes to a finite momentum ensures the remaining higher-energy magnons are degenerate with the continuum near the zone center. Experimentally, these conditions are also
likely to be satisfied by the zigzag ordered Na$_2$IrO$_3$ \cite{chun2015direct}, and spiral magnets $\alpha$-, $\beta$-, and
$\gamma$-Li$_2$IrO$_3$ \cite{PhysRevLett.113.197201,PhysRevB.90.205116,PhysRevB.93.195158}. This
picture is also consistent with recent indications that the magnetically
disordered phase observed at high pressure in $\beta$-Li$_2$IrO$_3$ \cite{PhysRevLett.114.077202} is driven primarily by large $\Gamma_1$
interactions \cite{PhysRevB.94.245127}. 

With this in mind, there are two general scenarios that can explain the
observed continuum excitations in $\alpha$-RuCl$_3$ and the iridates
$A_2$IrO$_3$. In the first scenario, which has been advanced by several studies, the excitations can be treated as free particles with a small number of flavours. Such excitations are weakly
interacting and have well-defined dispersions, but possess quantum numbers (e.g. $\Delta S_{\text{tot}} = \pm 1/2$) or topological properties inconsistent with the experimental neutron scattering selection rules (i.e. $\Delta S_{\text{tot}} = 0, \pm 1$). The appearance of the broad continuum in energy therefore results only from the fact that these fractional excitations must be created in multiples. If they could have been created individually, they would have represented long-lived and coherent quasiparticles with sharply peaked energies. 
This scenario indeed describes the Kitaev spin liquid, where the special symmetries of the Hamiltonian allow an
exact description in terms of two flavours of particles: non-interacting Majorana spinons and localized fluxes \cite{kitaev2006anyons}. Such excitations are long-lived, but belong to nontrivial topological sectors, and therefore cannot be created individually by any local operations. For the Kitaev spin liquid, the predicted continuum therefore represents coherent multiparticle excitations. 

In contrast, upon moving away from the pure Kitaev point, the relevant symmetries that protect the spinons and fluxes are lifted both by additional magnetic interactions and by spontaneous symmetry breaking of the magnetic order. This tends to
confine spinons into gauge neutral objects such as magnons \cite{PhysRevB.84.155121,PhysRevB.86.224417}. Despite this latter 
tendency, we have
argued that coherent magnons are unlikely to appear at large $\Gamma_1$ due to
the strong anharmonicity in the magnon Hamiltonian. While this leaves open the possibility that
nearly free Majorana spinons persist into the zigzag ordered phase in some regions of the Brillouin zone, a more
general scenario is that the observed continua represent fully incoherent excitations. In this second scenario, the excitations are not describable in terms of any type of free particles with small decay rates and well defined dispersions. The broad continua therefore reflects the absence of coherent quasiparticles altogether, rather than particular experimental selection rules related to fractionalization. 
At present, it is not clear which of
these scenarios applies to the iridates and $\alpha$-RuCl$_3$, although a key
role must be played by both the Kitaev $K_1$ and off-diagonal
couplings such as $\Gamma_1$. In any case, the study of these materials calls into question the stability of magnetic quasiparticles in the presence of strongly anisotropic interactions.

In summary, we have shown that all main features of the
magnetic excitations in $\alpha$-RuCl$_3$ \cite{banerjee2016neutron, banerjee2016proximate, ranneutron} are consistent with strongly
anisotropic interactions having signs and relative magnitudes in agreement with
{\it ab initio} predictions. The ferromagnetic Kitaev coupling ($K_1 <0$)
is supplemented by a
significant off-diagonal term ($\Gamma_1 >0$) that plays a crucial role in
establishing both the zigzag order and the observed continua. In the presence
of such interactions, the conventional magnon description breaks down even deep
in the ordered phase, due to strong coupling of the one-magnon and two-magnon
states. This effect is expected to persist over a large 
range of the phase diagram suggesting that the
observed continua in $\alpha$-RuCl$_3$ and the iridates $A_2$IrO$_3$ represent
a rich and general phenomenon extending beyond the Kitaev spin liquid. For this
class of strongly spin-orbital coupled magnets, the presence of complex and
frustrated anisotropic interactions leads naturally to dominant anharmonic
effects in the inelastic magnetic response. Fully describing the dynamics of
these and similar materials therefore represents a formidable challenge
that is likely to reveal aspects not found in conventional isotropic
magnets. \\
\textbf{Methods} \\
\textbf{Exact diagonalization.}
The neutron scattering intensity was computed via:
 \begin{align}
 \mathcal{I}(\mathbf{k},\omega) \propto& \  f^2(\mathbf{k})\int dt \ \sum_{\mu,\nu} (\delta_{\mu,\nu} - k_\mu k_\nu/k^2)\times \\ \nonumber & \ \times \sum_{i,j} \langle S_i^\mu (t) S_j^\nu (0)\rangle e^{-i\mathbf{k}\cdot (\mathbf{r}_i-\mathbf{r}_j)-i\omega t}\end{align}
where $f(\mathbf{k})$ is the atomic form factor of Ru$^{3+}$ from Ref.~\cite{Cromer:a04464}. ED calculations were performed using the Lanczos algorithm \cite{lanczos1950iteration}, on several 20- and 24-site clusters with periodic boundary conditions. Such periodic clusters are detailed in Supplementary Note 4. Excitations were computed using the continued fraction method \cite{RevModPhys.66.763}. Further details and additional results are presented in the supplemental material; these extensive calculations go beyond previous ED studies \cite{chaloupka2010kitaev,rau2014generic,PhysRevB.94.064435,katukuri2014kitaev,yadav2016spin}, which focused mainly on the static properties, or a limited portion of the phase diagram. ED results shown for the high-symmetry $\Gamma$, M, Y, X, and
$\Gamma^\prime$ points were averaged over all clusters. The 
ED $\mathbf{k}$-dependence of $\mathcal{I}(\mathbf{k},\omega)$, integrated over the 
energy windows $E = 1.3-2.3$, $5.5-8.5$, and $10.5+$ meV (Fig.~\ref{fig-spectralAFM}c and \ref{fig-spectralFM}c), was obtained from 
a single 24-site cluster respecting all symmetries of the model. 
The discrete ED spectra were Gaussian broadened by 0.5 meV, consistent with the width
of experimental features \cite{banerjee2016neutron}. The intensities were also
averaged over the same range of out-of-plane momentum as in the experiment \cite{banerjee2016neutron}.\\
\textbf{Linear spin-wave theory.}
LSWT results shown in Fig. 1 and 2 were obtained with the aid of SpinW \cite{toth2015linear}. Following the approach with the ED data, the discrete LSWT spectra were as well Gaussian broadened by 0.5 meV and the intensities were also
averaged over the same range of out-of-plane momentum as with ED and in the experiment \cite{banerjee2016neutron}.
\\
\textbf{Imaginary self-consistent Dyson equation approach.}
In order to calculate magnon decay rates $\gamma_{{\bf k},n}$, we have evaluated three-magnon interaction vertices by performing rotation to local reference frames of spins. The obtained value of the real-space interaction is quite large, about $\sim $ 3 meV. Next, the
Born approximation calculation of the decay rates results in unphysical divergencies~\cite{zhitomirsky2013colloquium}, thus the
self-energy $\Sigma_{{\bf k},n}$ needs to be regularized. We have used the so-called iDE approach: 
a self-consistent solution on the imaginary part of the Dyson's equation, $\Sigma_{{\bf k},n}(\epsilon_{\mathbf{k},n}+i\gamma_{{\bf k},n})\!=\! -i\gamma_{{\bf k},n}$, see Ref.~\cite{PhysRevB.94.140407}. We have obtained the 
regularized broadening for the magnon spectrum and have calculated the transverse part of the dynamical structure factor, shown
in Fig.~\ref{fig_ide}, by adding the calculated decay rates to experimental resolution of 0.25 meV. 
The spectral function is approximated as a Lorentzian. More technical details can be found in the Supplementary Note 3.

\textbf{Code availabilty} \\
Custom computer codes used in this study are
available from the corresponding author upon reasonable request. Documentation of
the codes is not available.

\textbf{Data availability} \\
Data is available from the corresponding author upon
reasonable request.

\textbf{Acknowledgements} \\
The authors acknowledge useful discussions with J. Chaloupka, A. Banerjee, S. E. Nagler, A. A. Tsirlin, R. Moessner, F. Pollmann, and M. Zhitomirsky. S.~M.~W. acknowledges support through an NSERC Canada
  Postdoctoral Fellowship. R.~V. and K.~R. acknowledge support by the
  Deutsche Forschungsgemeinschaft through grant SFB/TR 49. The work of P.~A.~M. and A.~L.~C. was supported by the U.S. Department of Energy, Office of Science, Basic Energy Sciences 
under Award No. DE-FG02-04ER46174.
  
\textbf{Author Contributions}\\
R.~V. and S.~M.~W. conceived the project. K.~R., S.~M.~W. and A.~H. performed and analyzed the ED calculations. P.~A.~M. and A.~L.~C. performed and analyzed the iDE results. All authors contributed equally to the manuscript.

\textbf{Competing interests}\\
The authors declare no competing financial interests.

\clearpage

\section{Figures}
\begin{figure}[h]
\includegraphics[width=0.8\linewidth]{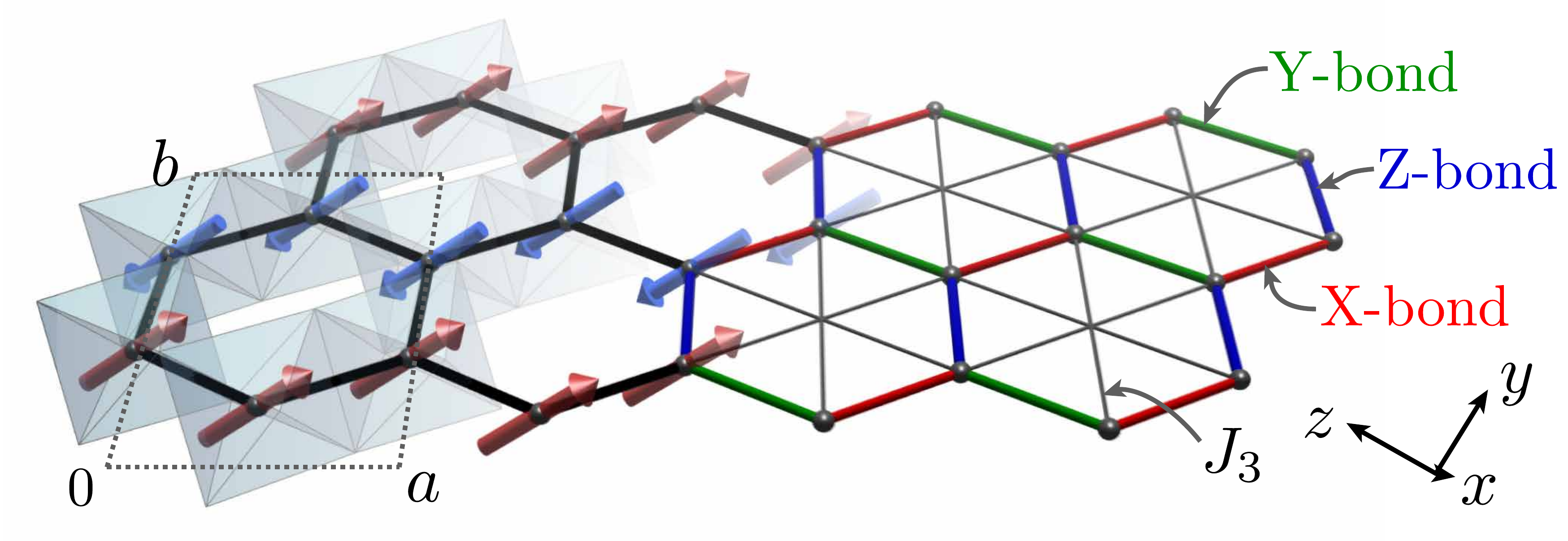}
\caption{\label{fig-1}\textsf{  \textbf{From material to model}. Within the honeycomb $ab$-layer of $\alpha$-RuCl$_3$ are illustrated the RuCl$_6$ octahedra, magnetic zigzag ordering pattern, and definition of the underlying magnetic interactions. Crystal axes are labelled with respect to the $C2/m$ structure.}}
\end{figure}
\begin{figure*}[h]
\includegraphics[width=\linewidth]{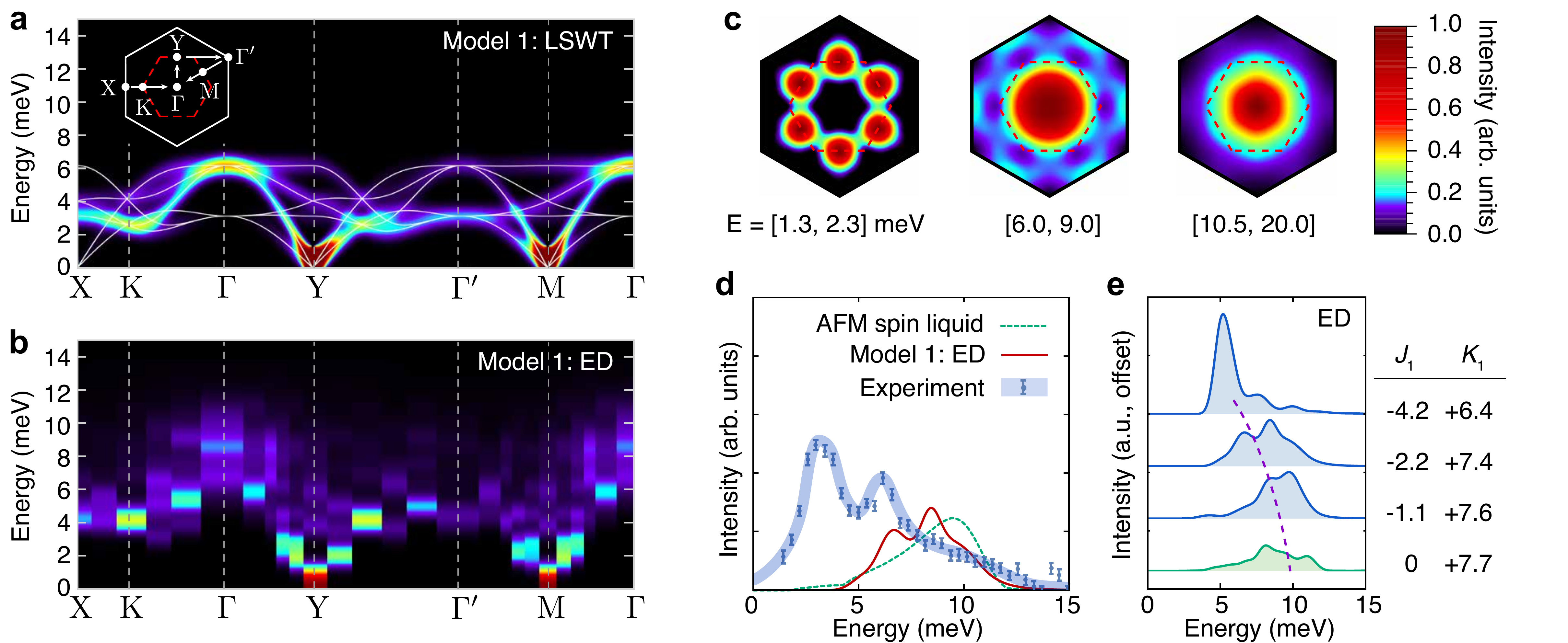}
\caption{ \textsf{ \label{fig-spectralAFM} \textbf{Neutron scattering intensity $\mathcal{I}(\mathbf{k},\omega)$ within the nnHK model.}  (\textsf{\textbf{a-c}}) Detailed results for Model 1 ($J_1 = -2.2, K_1 = +7.4$ meV): 
(\textsf{\textbf{a}}) $\mathcal{I}(\mathbf{k},\omega)$ computed via linear spin-wave theory (LSWT); results are averaged over the three zigzag ordering wavectors, parallel to the X-, Y-, and Z-bonds. Inset: Definition of Brillouin zone and high-symmetry $k$-points. (\textsf{\textbf{b}}) Exact diagonalization (ED) results, combining data from several 20- and 24-site periodic clusters (see Methods). (\textsf{\textbf{c}}) ED $\mathbf{k}$-dependence of $\mathcal{I}(\mathbf{k},\omega)$ integrated over the
indicated energies, as obtained from a single 24-site cluster respecting all symmetries of Eq.~\eqref{eqn-1} (see Methods). (\textsf{\textbf{d}}) Comparison of $\Gamma$-point intensities for the $K_1 = +7.7$ meV AFM spin liquid (exact results \cite{PhysRevB.92.115127,PhysRevLett.112.207203}), Model 1 (ED), and the experimental data for $\alpha$-RuCl$_3$ \cite{banerjee2016neutron}.
(\textsf{\textbf{e}}) Evolution of the ED $\Gamma$-point intensity with decreasing $|J_1/K_1|$, showing absence of low-energy intensity close to the $K_1>0$ spin liquid. The top three interaction sets correspond to zigzag order, while the bottom is the $K_1>0$ spin liquid. For all spectra, a Gaussian broadening of 0.5 meV has been applied.}}
\end{figure*}
\begin{figure*}[h]
\includegraphics[width=\linewidth]{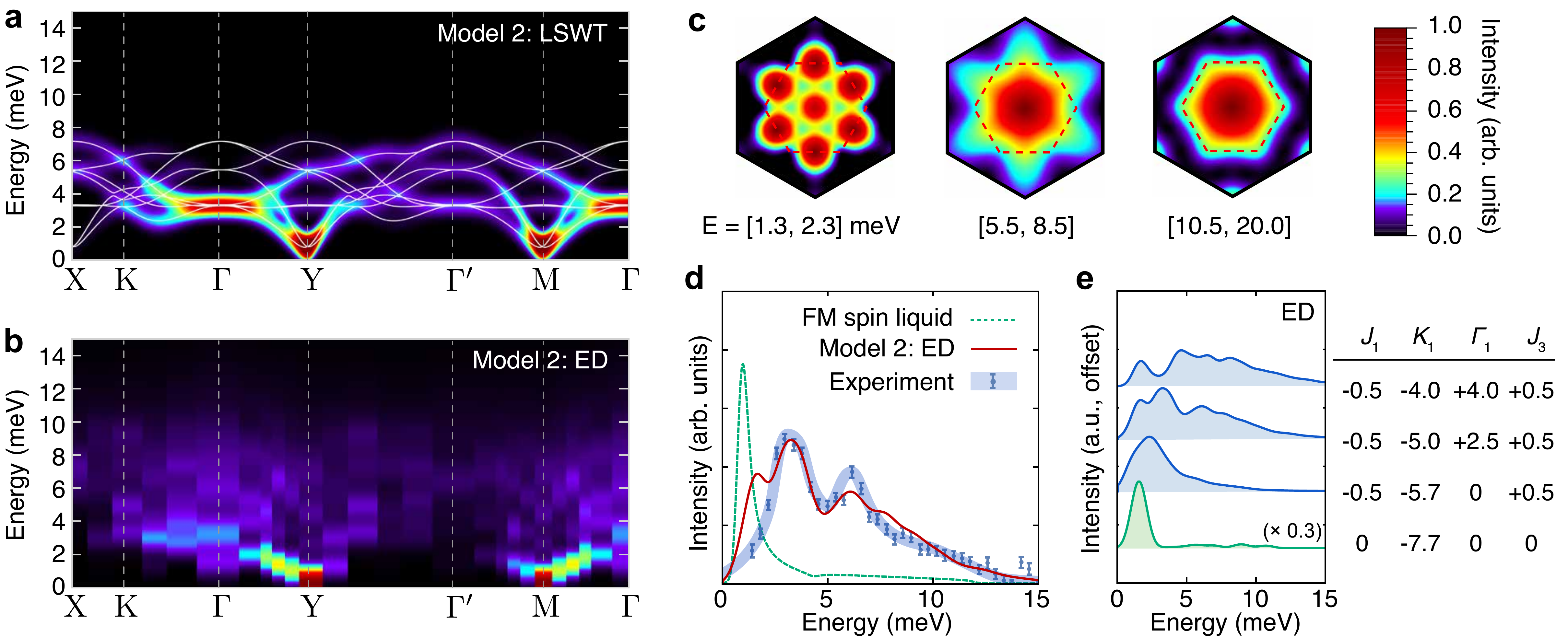}
\caption{\label{fig-spectralFM}  \textsf{\textbf{Neutron scattering intensity $\mathcal{I}(\mathbf{k},\omega)$ within the extended model.} (\textsf{\textbf{a-c}}) Detailed results for Model 2 ($J_1 = -0.5,K_1=-5.0,\Gamma_1=+2.5,J_3=+0.5$ meV): 
(\textsf{\textbf{a}}) $\mathcal{I}(\mathbf{k},\omega)$ computed via linear spin-wave theory (LSWT); results are averaged over the three zigzag domains with ordering wavectors parallel to the X-, Y-, and Z-bonds. (\textsf{\textbf{b}}) Exact diagonalization (ED) results, combining data from several 20- and 24-site periodic clusters (see Methods). (\textsf{\textbf{c}}) ED $\mathbf{k}$-dependence of $\mathcal{I}(\mathbf{k},\omega)$ integrated over the
indicated energies, as obtained from a single 24-site cluster respecting all symmetries of Eq.~\eqref{eqn-1} (see Methods). (\textsf{\textbf{d}}) Comparison of $\Gamma$-point intensities for the $K_1 = -7.7$ meV FM spin liquid (exact results \cite{PhysRevB.92.115127,PhysRevLett.112.207203}), Model 2 (ED), and the experimental data for $\alpha$-RuCl$_3$ \cite{banerjee2016neutron}.
(\textsf{\textbf{e}}) Evolution of the ED $\Gamma$-point intensity with decreasing $|\Gamma_1/K_1|$, showing significant broadening at finite $\Gamma_1$. The top three interaction sets correspond to zigzag order, while the bottom is the $K_1<0$ spin liquid. For all spectra, a Gaussian broadening of 0.5 meV has been applied.}}
\end{figure*}
\begin{figure}[h]
\includegraphics[width=0.7\linewidth]{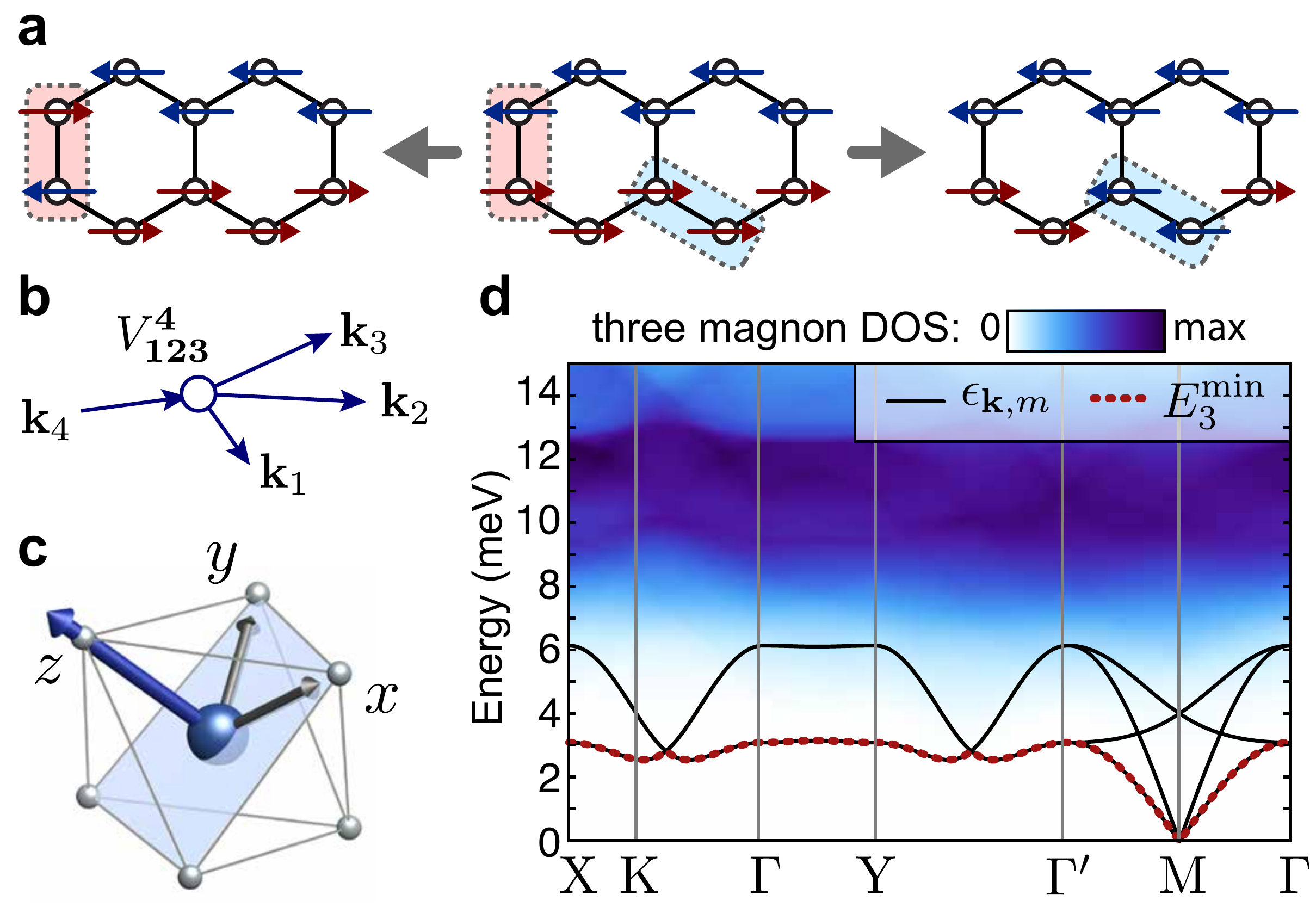}
\caption{\label{fig-AFMb}  \textsf{\textbf{Magnon decay channels for the nnHK model.} (\textsf{\textbf{a}}) Local
picture of quantum fluctuations away from zigzag order. The energy cost for the left process vanishes on approaching the spin liquid $|J_1/K_1| \rightarrow 0$.  (\textsf{\textbf{b}}) Momentum space picture for the corresponding fourth order decay process due to $\mathcal{H}_4$. (\textsf{\textbf{c}}) Ordered moment direction for Model 1 ($J_1 = -2.2, K_1 = +7.4$ meV), corresponding to the zigzag domain with ordering wavevector
$\mathbf{Q} = $ Y. (\textsf{\textbf{d}}) LSWT dispersions
$\epsilon_{\mathbf{k},m}$, and 3-magnon density of states (DOS) for Model 1 for the same zigzag domain as (\textsf{\textbf{c}}). The dashed line indicates the bottom of the 
three-magnon continuum ($E_3^{\text{min}}$), which is coincident with the lowest magnon band.} }\end{figure}

\begin{figure}[h]
\includegraphics[width=0.7\linewidth]{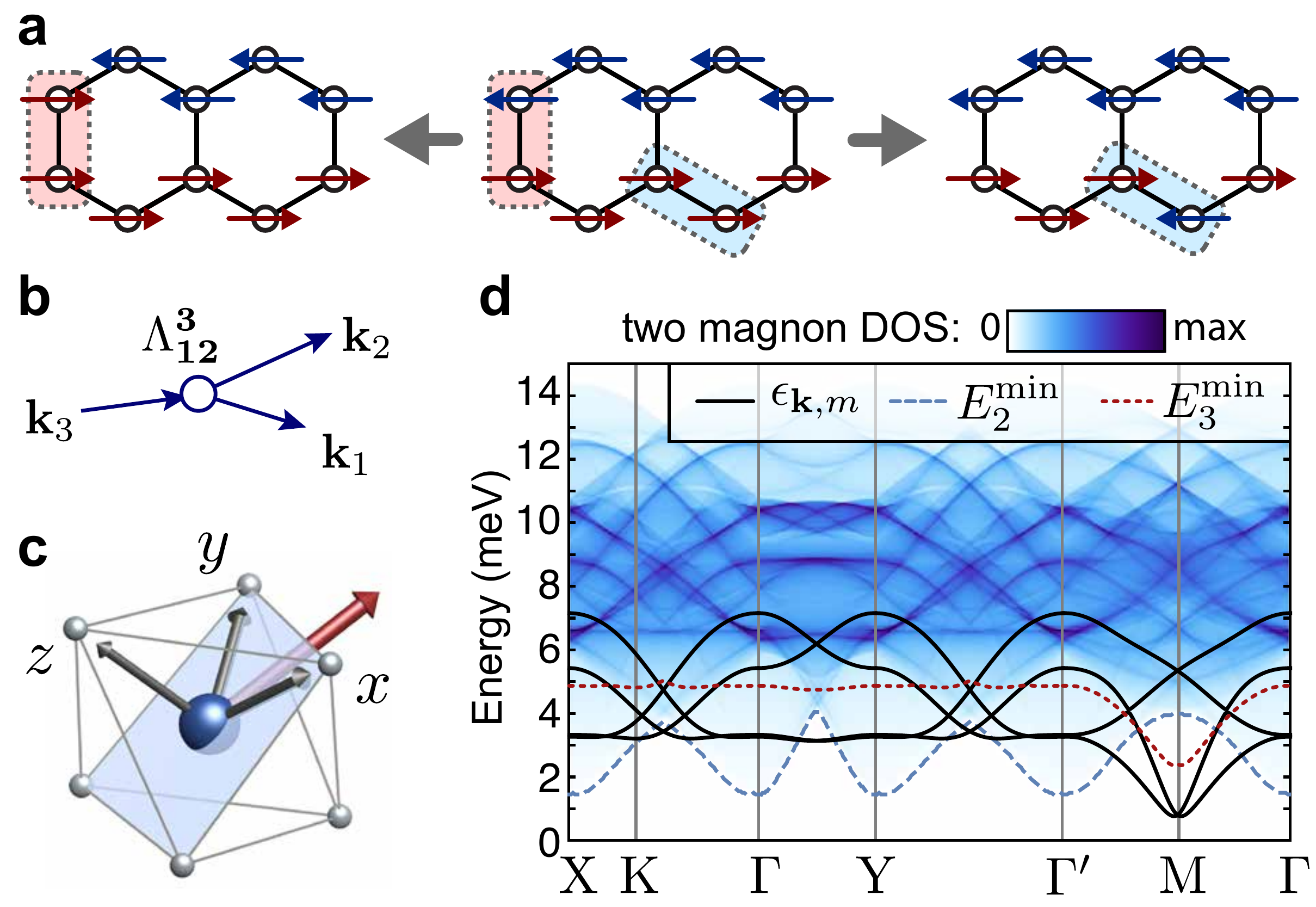}
\caption{\label{fig-FMb} \textsf{ \textbf{Magnon decay channels for the extended model.} (\textsf{\textbf{a}}) Local picture of additional quantum fluctuations away from zigzag order induced by $\Gamma_1$ interactions. (\textsf{\textbf{b}}) Momentum space picture of the third order decay process $\mathcal{H}_3$. (\textsf{\textbf{c}}) Ordered moment direction for Model 2  ($J_1 = -0.5,K_1=-5.0,\Gamma_1=+2.5,J_3=+0.5$ meV) with zigzag ordering wavevector $\mathbf{Q} = $ Y, parallel to the Z-bond. (\textsf{\textbf{d}}) LSWT dispersions $\epsilon_{\mathbf{k},m}$, and 2-magnon density of states (DOS) for Model 2 with $\mathbf{Q} = $ Y. Dashed lines indicate the bottom of the two- and three-magnon continuum ($E_2^{\text{min}}(\mathbf{k})$ and $E_3^{\text{min}}(\mathbf{k})$, respectively).}}
\end{figure}

\begin{figure*}[h]
\includegraphics[width=\linewidth]{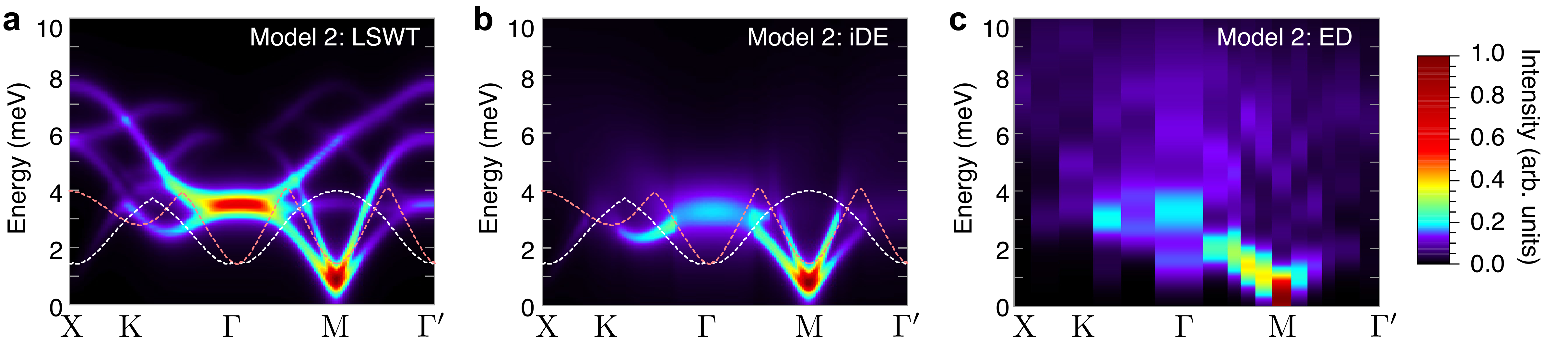}
\caption{\label{fig_ide} \textsf{ \textbf{Effects of two-magnon decays in $\mathcal{I}(\mathbf{k},\omega)$ for extended model}. Results are shown for Model 2 computed via (\textsf{\textbf{a}}) linear spin-wave theory (LSWT), (\textsf{\textbf{b}}) self-consistent imaginary Dyson equation (iDE) approach, and (\textsf{\textbf{c}}) exact diagonalization (ED). Results in  (\textsf{\textbf{a}}) and (\textsf{\textbf{b}}) are averaged over the different zigzag domains. The white and pink dashed lines indicate the bottom of the two-magnon continuum, $E_2^{\text{min}}(\mathbf{k})$ for the different zigzag domains. In the iDE results, the effects of two-magnon decays strongly broadens any magnon bands overlapping with the two-magnon continuum. }}
\end{figure*}
\clearpage



\section*{Supplementary material}

# Supplemental Material for: Breakdown of Magnons in a Strongly Spin-Orbital Coupled Magnet

Stephen M. Winter,[1],[*] Kira Riedl,[1] Pavel A. Maksimov,[2] Alexander L. Chernyshev,[2] Andreas Honecker,[3] and Roser Valentí[1]

[1]*Institut für Theoretische Physik, Goethe-Universität Frankfurt, Max-von-Laue-Strasse 1, 60438 Frankfurt am Main, Germany*
[2]*Department of Physics and Astronomy, University of California, Irvine, California 92697, USA*
[3]*Laboratoire de Physique Théorique et Modélisation, CNRS UMR 8089, Université de Cergy-Pontoise, 95302 Cergy-Pontoise Cedex, France*

(Dated: October 31, 2017)

# SUPPLEMENTARY NOTE 1:
# PHASE DIAGRAMS OF THE EXTENDED $(J_1, K_1, \Gamma_1, J_3)$ MODEL

In this section, we compute the phase diagram for the extended $(J_1, K_1, \Gamma_1, J_3)$ spin Hamiltonian on the honeycomb lattice discussed in the main text, and given by:

$$\mathcal{H} = \sum_{\langle ij \rangle} J_1\, \mathbf{S}_i \cdot \mathbf{S}_j + K_1\, S_i^\gamma S_j^\gamma + \Gamma_1\, (S_i^\alpha S_j^\beta + S_i^\beta S_j^\alpha) + \sum_{\langle\langle\langle ij \rangle\rangle\rangle} J_3\, \mathbf{S}_i \cdot \mathbf{S}_j \qquad (1)$$

where $\{\alpha, \beta, \gamma\} = \{y, z, x\}$ for the $X$ bonds, $\{z, x, y\}$ for the $Y$ bonds, and $\{x, y, z\}$ for the $Z$ bonds shown in Fig. 1 of the main text. In order to capture the effect of local quantum fluctuations on the state energies, we computed second-order energy corrections to the classical ordered state energies, extending the method of Supplementary Ref. [1] to include finite $\Gamma_1$ and $J_3$. As discussed in Supplementary Refs. [1] and [2], an upper bound on the energies per site of the Kitaev spin-liquid states can be estimated from $\mathcal{E}_\mathrm{KIT} = \pm\frac{3}{2}(J_1 + K_1)\langle S_i^\gamma S_j^\gamma \rangle$, where $\langle S_i^\gamma S_j^\gamma \rangle \approx 0.131$ is the analytical result for the first neighbour correlations at the pure $K_1 > 0$ or $K_1 < 0$ Kitaev points [3]. Comparison of $\mathcal{E}_\mathrm{KIT}$ with the second-order corrected energies of the ordered states has been shown to reliably predict the position of the phase boundaries, which agree with the results of exact diagonalization (ED) [1, 2].

We show in Supplementary Figure 1a,b the phase diagram associated with Supplementary Eq. (1), parameterizing $J_1 = \cos\phi \sin\theta$, $K_1 = \sin\phi \sin\theta$, and $\Gamma_1 = \cos\theta$ as in Supplementary Ref. [4], with $J_3 = 0$ (Supplementary Figure 1a) and $J_3 > 0$ (Supplementary Figure 1b). The present results may be compared directly with the ED results of Supplementary Ref. [4] for $J_3 = 0$. The extended model of Supplementary Eq. (1) exhibits six ordered phases, which have been identified in various previous works [1–10]: FM = collinear ferromagnetic order, AFM = collinear Néel antiferromagnetic order, ST = collinear stripy order, ZZ = collinear zigzag order, 120 = noncollinear 120° order, and IC = incommensurate spiral order. In the extended model, the classical energies per site are given by:

$$\mathcal{E}_\mathrm{FM} = \frac{1}{8}(3J_1 + K_1 - \Gamma_1 + 3J_3) \qquad (2)$$

$$\mathcal{E}_\mathrm{AFM} = -\frac{1}{8}(3J_1 + K_1 + 2\Gamma_1 + 3J_3) \qquad (3)$$

$$\mathcal{E}_{120} = -\frac{1}{8}(K_1 + 2\Gamma_1) \qquad (4)$$

$$\mathcal{E}_\mathrm{ST} = \frac{1}{16}\left(-2J_1 + \Gamma_1 + 6J_3 - \sqrt{9\Gamma_1^2 - 4\Gamma_1 K_1 + 4K_1^2}\right) \qquad (5)$$

$$\mathcal{E}_\mathrm{ZZ} = \frac{1}{16}\left(2J_1 - \Gamma_1 - 6J_3 - \sqrt{9\Gamma_1^2 - 4\Gamma_1 K_1 + 4K_1^2}\right) \qquad (6)$$

$$\mathcal{E}_\mathrm{IC} = \frac{1}{2}\left(K_1 - \Gamma_1 - \sqrt{8\Gamma_1^2 + K_1^2}\right). \qquad (7)$$

The full expressions with second-order corrections are lengthy, and therefore omitted here for brevity. However, the general trends are already apparent from Supplementary Eq. (2)-(7).

Of particular interest are the regions of stability of the zigzag order, observed experimentally in the honeycomb materials $\alpha$-RuCl$_3$ and Na$_2$IrO$_3$, as well as the extent of the $K_1 > 0$

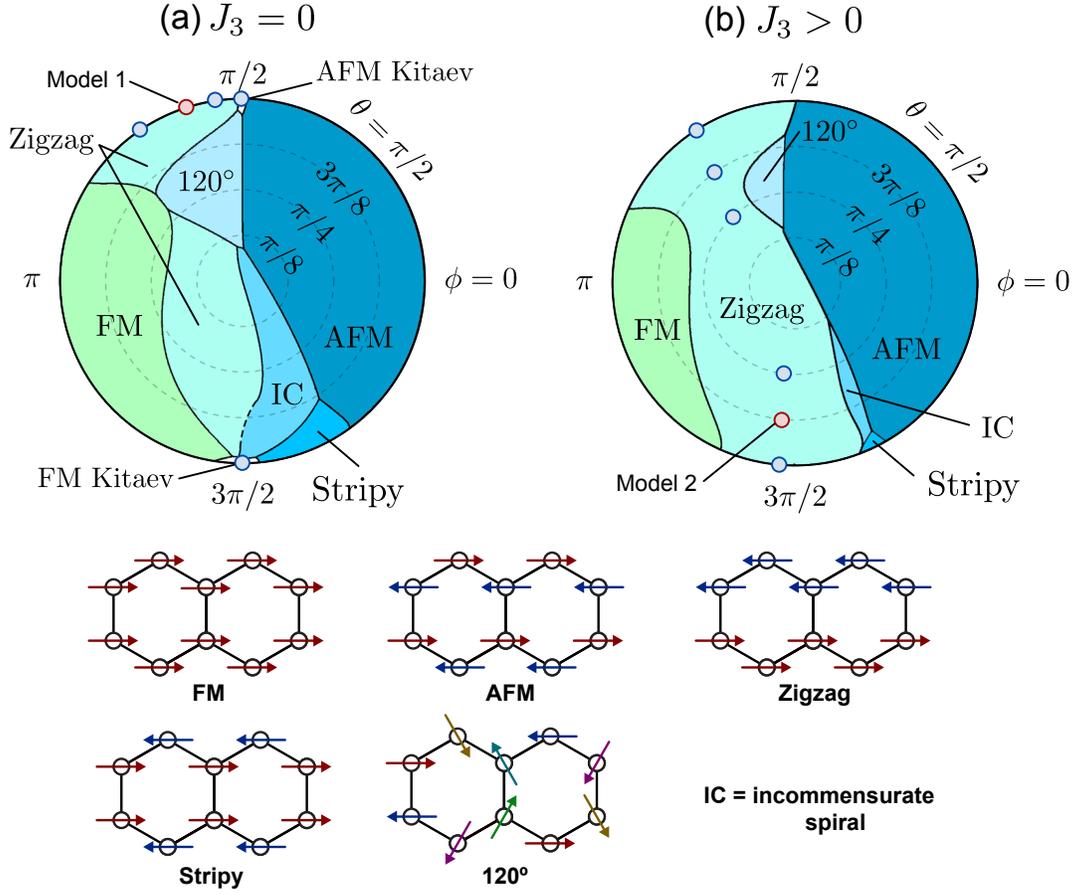

SUPPLEMENTARY FIGURE 1. **Phase diagrams for the extended model.** The phases were computed as described in the text. (a) With $J_3 = 0$. The dashed line between the incommensurate and zigzag phases indicates a region where $\mathcal{E}_{\text{ZZ}} \approx \mathcal{E}_{\text{IC}}$ and $\partial\mathcal{E}_{\text{ZZ}}/\partial J \approx \partial\mathcal{E}_{\text{IC}}/\partial J$, reducing the accuracy of the boundary line. Results here can be compared directly with the ED results of Supplementary Ref. [4]. (b) With a relatively small $J_3/\sqrt{J_1^2 + K_1^2 + \Gamma_1^2} = +0.088$ consistent with the magnitude in the studied models. Small circles indicate the considered parameters. For the incommensurate spiral state, we did not consider quantum modifications to the ordering wavevector. The models studied in Supplementary Note 5 are highlighted by blue and red points.

and $K_1 < 0$ Kitaev spin-liquids. For $J_3 = 0$, there are two zigzag regions, appearing at $(J_1 < 0, K_1 > 0, \Gamma_1 \approx 0)$, and $(J_1 \approx 0, K_1 < 0, \Gamma_1 > 0)$. The addition of a small finite $J_3$ uniquely stabilizes the zigzag and Néel states, linking the two zigzag regions, and suppressing the spin-liquid and other ordered phases. The existence of such $J_3$ coupling has been indicated for both $\alpha$-RuCl$_3$ [8, 10] and Na$_2$IrO$_3$ [8, 9, 11]. This fact significantly complicates the identification of the magnetic interactions in these materials from investigations of the static properties alone, since the zigzag phase is stable over a very wide region of the phase diagram. The stability region for the $K_1 > 0$ and $K_1 < 0$ Kitaev spin-liquid states estimated from the second-order state energies is shown in Supplementary Table 1.

The relevant energy scale for considering the stability of the spin-liquid is the energy gap for spin-excitations at the pure Kitaev points, given by the two-flux gap $\Delta \approx 0.065|K_1|$

|  | $K_1 = +1$ | $K_1 = -1$ |
| --- | --- | --- |
| $\Gamma_1 = 0, J_3 = 0:$ | $-0.023 < J_1 < +0.025$ | $-0.160 < J_1 < +0.095$ |
| $J_1 = 0, J_3 = 0, \Gamma_1 > 0:$ | $\Gamma_1 < +0.140$ | $\Gamma_1 < +0.054$ |
| $J_1 = 0, \Gamma_1 = 0, J_3 > 0:$ | $J_3 < +0.041$ | $J_3 < +0.053$ |

SUPPLEMENTARY TABLE 1. **Stability region for Kitaev spin-liquid states.** The regions were estimated from second-order state energies.

[12, 13]. Indeed, at the pure Kitaev points, the above estimates for the state energies suggest that the Kitaev state is stabilized with respect to adjacent magnetic orders by $0.072|K_1| \approx \Delta$ per site. In this sense, any perturbation on the scale of $\Delta$ has the potential to destabilize the spin-liquid. This explains why the Kitaev spin-liquid occupies a relatively small region of the phase diagram.

## SUPPLEMENTARY NOTE 2:
## REVIEW OF *AB INITIO* STUDIES OF $\alpha$-RuCl$_3$

In this section, we briefly review previous *ab initio* studies of the magnetic interactions in $\alpha$-RuCl$_3$. As discussed by some of the present authors in Supplementary Ref. [8], estimation of such interactions from first principles calculations are complicated by several factors:

- The layered structure of $\alpha$-RuCl$_3$ allows for significant stacking defects in the crystal structure, which have complicated structural solution. Very early studies indicated a trigonal space group $P3_112$ [14, 15], with $\angle$Ru-Cl-Ru $\sim 88°$. More recent detailed reanalysis of the structure suggested it to be monoclinic $C2/m$ [16, 17] at low temperatures, with $\angle$Ru-Cl-Ru $\sim 94°$. Furthermore, there is now increasing evidence that $\alpha$-RuCl$_3$ exhibits a structural phase transition near $T \sim 150$ K [18, 19], which may be analogous to the $C2/m \to R\bar{3}$ transition of CrCl$_3$ [20].

  Given that the magnetic interactions are highly sensitive to the local geometry of the RuCl$_6$ octahedra, accurate estimation of their values has historically been complicated by the uncertainty in the crystal structure. This had led to a variety of reported interactions for $\alpha$-RuCl$_3$, summarized in Supplementary Table 2.

- The low symmetry of the real crystal structures allows many independent terms to appear in the spin Hamiltonian. For example, up to 30 parameters are required to fully define the interactions up to third nearest neighbours [8]. As in the main text, most previous works have treated only a selection of such parameters, therefore assuming the interactions to be of higher symmetry than required by the crystal structure. This corresponds to an effective averaging of the interactions, which is likely to produce variations in the computed magnitudes of each parameter across different computational methods.

- For $\alpha$-RuCl$_3$, the underlying energy scales (Hund's coupling, spin-orbit coupling, and crystal-field splitting, for example) are all of similar magnitude, which makes the computed interactions highly sensitive to fine details (such as structure, and choice

| Structure | Method | sgn($K_1$) | $J_1/K_1$ | $\Gamma_1/K_1$ | $J_3/K_1$ | Ref. |
|---|---|---|---|---|---|---|
| $P3_112$ [14] | DFT+2OPT | $K_1 > 0$ | $-0.7$ | $+0.7$ | $+0.02$ | [21] |
| $P3_112$ [15] | DFT+ED | $K_1 > 0$ | $-0.7$ | $+1.1$ | $+0.3$ | [8] |
| $P3_112$ [14] | QC | $K_1 < 0$ | $+0.4$ | $-0.8$ | $-$ | [10] |
| $C2/m$ [23] | DFT+2OPT | $K_1 < 0$ | $+0.1$ | $-0.5$ | $-$ | [23] |
| $C2/m$ [16] | DFT+ED | $K_1 < 0$ | $+0.25$ | $-1.0$ | $-0.4$ | [8] |
| $C2/m$ [16] | QC | $K_1 < 0$ | $-0.1$ | $-0.25$ | $-$ | [10] |
| $C2/m$ [17] | DFT | $K_1 < 0$ | $+0.15$ | $-0.35$ | $-0.1$ | [24] |
| $C2/m$ [17] | DFT+2OPT | $K_1 < 0$ | $\sim 0$ | $-0.55$ | $-$ | [25] |

SUPPLEMENTARY TABLE 2. **Summary of previous interaction parameters for $\alpha$-RuCl$_3$.** 2OPT = second order perturbation theory, ED = exact diagonalization, QC = quantum chemistry, and DFT = density functional theory. Since the magnitude of interactions may generally differ along the $X$, $Y$, and $Z$ bonds, we present bond-averaged values.

of Coulomb parameters). This provides additional variations in computed interaction magnitudes appearing in the literature.

Despite these complications, estimates of the interaction parameters for $\alpha$-RuCl$_3$ have broadly agreed across many different methods, when similar crystal structures are taken as input. A summary of previous *ab initio* calculations for such interactions is shown in Supplementary Table 2.

*Ab initio* studies based on the early $P3_112$ structures [14, 15] of $\alpha$-RuCl$_3$ suggested antiferromagnetic $K_1 > 0$, with ferromagnetic $J_1 < 0$ and antiferromagnetic $\Gamma_1 > 0$ of similar magnitude [8, 21]. Microscopically, the positive $K_1$ arises from large direct hopping between Ru metal sites [4], which has a dramatic effect for the short Ru-Ru distances in the $P3_112$ structure. It should be emphasized that this hopping also generates large $\Gamma_1$ interactions, such that antiferromagnetic $K_1 > 0$ must always be accompanied by large $\Gamma_1$ in real materials. For this reason, pure $(J_1, K_1)$ nnHK interactions suggested in Supplementary Ref. [22], based on spin-wave fitting, are inconsistent with $K_1 > 0$, from a microscopic perspective.

Later *ab initio* studies based on the (more accurate) $C2/m$ structures [16, 17] have instead suggested $K_1 < 0$, $J_1 \sim 0$, and $\Gamma_1 > 0$ [8, 10, 23–25]. For reasons suggested above, there has been a relatively large spread of computed values obtained from different computational methods. However, if we do not favour any particular method, we might expect an appropriate starting point for analysis to appear at the average values over all studies:

$$(J_1/K_1)_{\text{avg}} \sim +0.08 \quad , \quad (\Gamma_1/K_1)_{\text{avg}} \sim -0.5 \tag{8}$$

and $K_1 < 0$. As noted in the main text, excellent agreement between the ED and experimental neutron spectra is obtained for Model 2, with $(J_1/K_1) = +0.1$, and $(\Gamma_1/K_1) = -0.5$, which is completely consistent with the range of *ab initio* values appearing in the literature.

5

# SUPPLEMENTARY NOTE 3:
# THREE-MAGNON COUPLINGS, DECAY KINEMATICS AND RATES, AND DYNAMICAL STRUCTURE FACTOR

**Three-Magnon Coupling**

In the main text we write: "a large decay rate is ensured by the following three conditions: large anisotropic interactions, deviation of the ordered moments away from the high-symmetry axes, and strong overlap of the one-magnon states with the multi-magnon continuum." Below we elaborate on these conditions for strong magnon decays.

The Hamiltonian for $\alpha$-RuCl$_3$ is given by Supplementary Eq. (1). The diagonalization of the Hamiltonian requires rotation to the local reference frames of spins. Taking the zigzag ordering wavevector $\mathbf{Q} = Y$, the ordered moment is expected to lie in the crystallographic $ac$-plane provided $\Gamma_1 > 0$, as for Models 1 and 2 of the main text, see also Supplementary Figure 2. This plane also contains the cubic $z$-axis. Thus, defining $\theta$ as the angle between the cubic $z$-axis and the ordered moment direction $\tilde{z}$, the spin operators can be rotated into the ordered coordinate frame via a rotation matrix:

$$\mathbf{R}_\theta = \begin{pmatrix} \cos^2\left(\frac{\theta}{2}\right) & -\sin^2\left(\frac{\theta}{2}\right) & \frac{1}{\sqrt{2}}\sin\theta \\ -\sin^2\left(\frac{\theta}{2}\right) & \cos^2\left(\frac{\theta}{2}\right) & \frac{1}{\sqrt{2}}\sin\theta \\ -\frac{1}{\sqrt{2}}\sin\theta & -\frac{1}{\sqrt{2}}\sin\theta & \cos\theta \end{pmatrix}, \tag{9}$$

where $\theta$ is obtained by a minimization of the classical energy and depends on $K_1/\Gamma_1$. Then the Hamiltonian is

$$\mathcal{H} = \sum_{\langle ij \rangle} \tilde{\mathbf{S}}_i \cdot \tilde{\mathbf{J}}_{ij} \cdot \tilde{\mathbf{S}}_j, \tag{10}$$

where $\tilde{\mathbf{J}}_{ij}$ is the bond-dependent exchange matrix in the local spin basis.

The coupling of the one- and two-magnon excitations is generated by the off-diagonal ("odd") terms that contain $\tilde{S}^z_i \tilde{S}^x_j$ and $\tilde{S}^z_i \tilde{S}^y_j$ in the local reference frame. Extracting such terms explicitly from the exchange matrix (10) gives

$$\mathcal{H}_{\text{odd}} = \sum_{\langle ij \rangle} \left( J^{\text{odd}}_{ij,x} S^x_i S^z_j + J^{\text{odd}}_{ij,y} S^y_i S^z_j + i \leftrightarrow j \right), \tag{11}$$

For the zigzag structure with $\mathbf{Q} = Y$, there are five distinct bonds with respect to the values of $J^{\text{odd}}_{ij,x}$, $J^{\text{odd}}_{ij,y}$. Thus, for the $X$ and $Y$ bonds, $J^{\text{odd}}_{ij,x(y)} = \pm A_\pm$, with

$$A_\pm = \frac{\Gamma_1}{2}(\cos 2\theta \mp \cos\theta) + \frac{K_1}{2\sqrt{2}}(\cos\theta \pm 1)\sin\theta, \tag{12}$$

and for the $Z$ bonds, $J^{\text{odd}}_{ij,x(y)} = B$, where

$$B = \frac{1}{\sqrt{2}}(\Gamma_1 - K_1)\cos\theta \sin\theta. \tag{13}$$

Thus, magnon decay vertices scale as $\sim (A_\pm, B)$. It is immediately apparent that the exchange terms do not contribute to the three-magnon coupling because of the collinear spin arrangement in a zigzag structure [26].

6

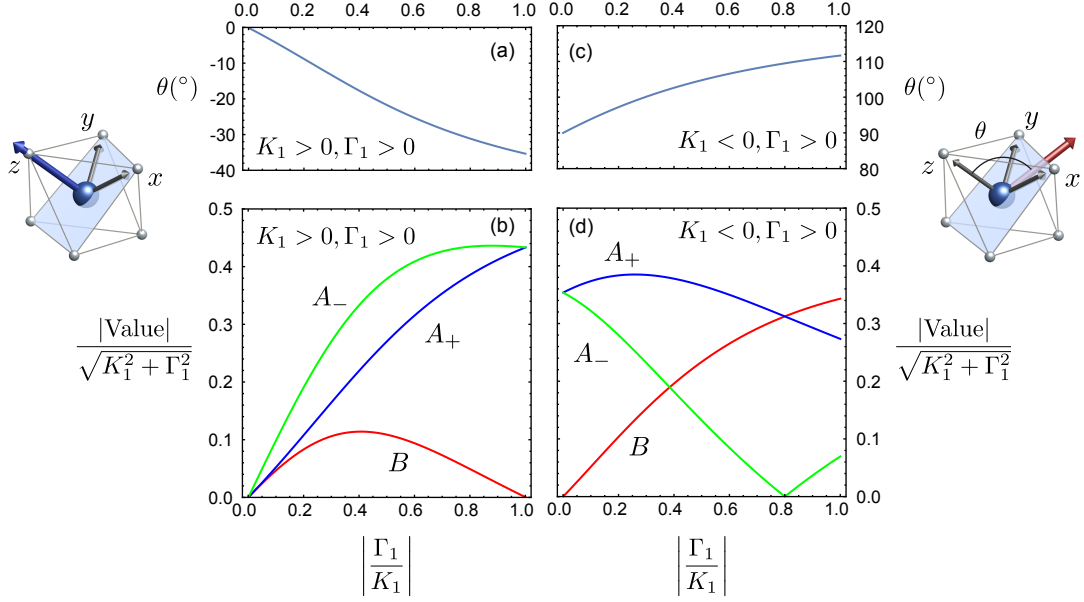

SUPPLEMENTARY FIGURE 2. **Off-diagonal exchange terms.** This figure shows the evolution of the direction of the ordered moment and of the off-diagonal ("odd") exchange terms (11) vs $|\Gamma_1/K_1|$. For (a) and (b), $K_1 > 0, \Gamma_1 > 0$, and for (c) and (d), $K_1 < 0, \Gamma_1 > 0$. In (b) and (d), terms contributing to the two-magnon decay vertex are shown. Their values are normalized to $\sqrt{K_1^2 + \Gamma_1^2}$, a rough magnitude of the single-magnon dispersion.

For the case of antiferromagnetic Kitaev coupling $K_1 > 0$, the ordered moments tend to align along the cubic z-axis, thus selecting $\theta \sim 0$ [27]. In Supplementary Figure 2a, we show the evolution of $\theta$ with $|\Gamma_1/K_1|$. For $\Gamma_1 = 0$, as in Model 1 of the main text, the ordered moment is exactly along the high-symmetry cubic z-axis ($\theta = 0$), and the off-diagonal $A_\pm$ and $B$ terms vanish according to Supplementary Eqs. (12) and (13) above. As such, the decay of one-magnon excitations into the two-magnon continuum is forbidden in the zigzag phase for the pure Heisenberg and Kitaev interactions. In this case, all spin interactions in the local spin basis appear in the form of $\tilde{S}_i^x \tilde{S}_i^x$, $\tilde{S}_i^y \tilde{S}_i^y$, or $\tilde{S}_i^z \tilde{S}_i^z$, which contribute only to the terms of even order in magnon operators. As discussed in Supplementary Ref. [27], this ordered moment direction is stable with regard to small $\Gamma_1 > 0$, which shifts the moments to $\theta < 0$, with $\theta \propto |\Gamma_1/K_1|$. In this case, the average magnitude of the two magnon decay vertex is expected to scale linearly with the off-diagonal $\Gamma_1$ couplings for small $\Gamma_1$. That is $\Lambda \sim (A_\pm, B) \propto |\Gamma_1/K_1|$. The evolution of $(A_\pm, B)$ with $|\Gamma_1/K_1|$ is shown in Supplementary Figure 2b, normalized to the overall magnitude of interactions $\epsilon_{\mathbf{k},m} \sim \sqrt{K_1^2 + \Gamma_1^2}$, which sets the scale of the one-magnon bandwidth. One can see that the decay terms become significant compared to the one magnon dispersion for large $\Gamma_1$.

For the ferromagnetic Kitaev coupling $K_1 < 0$, and $\Gamma_1 > 0$ (as in Model 2 of the main text), the situation is somewhat different. Finite $\Gamma_1$ terms rotate the ordered moment away from the cubic z-axis by an angle $\theta \gtrsim 90°$ for any value of $|\Gamma_1/K_1| \gtrsim 0.05$ (see Supplementary Ref. [27]). Thus, the moments lie close to the cubic $\hat{x} + \hat{y}$ direction. In the rotated coordinate frame, both $\Gamma_1$ and Kitaev interactions contribute to the "odd" terms in (11), and thus induce two-magnon decays. Because of that, $A_\pm$ or $B$ are always large and scale with the magnitude of $K_1$ and $\Gamma_1$, as shown in Supplementary Figure 2d. For that reason, in this

region of the phase diagram, the magnon decay vertex $\Lambda \sim (A_\pm, B) \sim (K_1, \Gamma_1)$ is always expected to be of the order of the one-magnon bandwidth.

The Holstein-Primakoff bosonization of (11) yields the three-boson Hamiltonian $\mathcal{H}_3$

$$\mathcal{H}_3 = \frac{1}{z} \sum_{\langle ij \rangle} \widetilde{J}_{ij}^{\text{odd}} \left( a_i^\dagger a_j^\dagger a_j + \text{H.c.} + i \leftrightarrow j \right), \tag{14}$$

where we extracted the coordination number $z=3$. Here,

$$\widetilde{J}_{ij}^{\text{odd}} \equiv 3\sqrt{\frac{S}{2}} \left( J_{ij,x}^{\text{odd}} + i J_{ij,y}^{\text{odd}} \right). \tag{15}$$

Using (12) and (13) for $S=1/2$ and the parameters of Model 2, the three-magnon coupling for bonds $X$, $Y$ and $Z$ are

$$\left| \widetilde{J}_{X(Y)}^{\text{odd}} \right| = 3.35 \text{meV}, \quad \left| \widetilde{J}_Z^{\text{odd}} \right| = 2.78 \text{meV}. \tag{16}$$

The quantities in (16) can be referred to as the real-space three-magnon vertices. For Model 2, their strength relative to the full magnon bandwidth $W \approx 7$ meV is $\approx 0.4 - 0.5$. Such a strong anharmonic coupling is a precursor of strong magnon decays.

**Decay Kinematics**

The consideration above ensures that the amplitude of the anharmonic magnon-coupling terms is significant in case of Model 2 as well as for a wider part of the phase diagram. However, the effect of this coupling on the single-magnon states depends crucially on the availability of the two-magnon states for decays. A strong overlap of one-magnon states with the two-magnon continuum for the high-energy magnon branches is virtually guaranteed by the presence of the low-energy branches, as is exemplified favour of a significant overlap of such kind for all branches of the spectrum, including the lowest one. That argument, together with a strong three-magnon coupling, in turn guarantees substantial magnon line broadenings.

The situation of interest is illustrated in a sketch in Supplementary Figure 3. For a simple magnon branch with a Goldstone mode at $\mathbf{k} = 0$ one can see that the bottom of the two-magnon continuum is precisely degenerate with the one-magnon dispersion. Moreover, the density of the two-magnon states vanishes at the one-magnon energy, rendering decays impossible [26]. If a gap $\Delta$ is introduced in the magnon energy at $\mathbf{k} = 0$, the two-magnon continuum will still have a minimum at the same $\Gamma$ point, but will be gapped with the energy $2\Delta$.

Next is the case of the gapless modes occurring at finite $\pm \mathbf{k}_0$, keeping an overall shape of the dispersion as simple as possible. It is obvious that now the two magnons with the total momentum $\mathbf{q} = 0$ have the bottom of their continuum at $\text{Min}(\epsilon_\mathbf{k} + \epsilon_{-\mathbf{k}}) = 0$. Thus, the two-magnon continuum must be below the one magnon states in an extended vicinity of $\mathbf{q} = 0$. It is also obvious that this argument is robust against a finite gap at the (pseudo-) Goldstone point as the overlap of the magnon branch with the continuum survives for a finite $\Delta$.

We mention that such a situation is not uncommon and describes almost exactly the case of a spiral antiferromagnet, which has an ordering vector at finite momentum, see

[26]. Another generic case are the spin-ladder and 2D valence-bond-like antiferromagnets having band minima at a finite $\mathbf{Q}$. The decay kinematic conditions and magnon decays are well-documented for them, both theoretically and experimentally, see [26] and [28, 29] for a recent realization.

While, in general, one can formulate a complete list of checks for decay conditions for an arbitrary form of the magnon spectrum, see [26], the current consideration provides a clear and intuitive picture of why magnon decays must occur in Model 2 and also more broadly.

**Decay Rates**

Using standard diagrammatic rules with the decay terms, one can straightforwardly obtain magnon self-energies in the one-loop (Born) and on-shell approximations, $\Sigma_{\mu\mathbf{k}}(\omega) \rightarrow \Sigma_{\mu\mathbf{k}}(\varepsilon_{\mu\mathbf{k}})$, both strictly within the $1/S$ expansion [26]. This neglects such effects as more complicated spectrum renormalizations and spectral weight redistribution away from the quasiparticle pole. One can argue that the real part of the self-energy should be neglected altogether as the renormalization of magnon energies is already built in by the choice of the model parameters. In practice, the LSWT parameters are indeed often chosen to best fit the observed experimental and/or numerical bands.

Altogether, this leaves us with the only remaining and yet the most important and physically distinct effect of the anharmonic terms: magnon decays. Thus, the self-energy of the magnon branch $\mu$ is $\Sigma_{\mu\mathbf{k}}(\omega) \rightarrow -i\gamma_{\mathbf{k}}^{\mu}$, and the decay rate in the Born approximation is

$$\gamma_{\mathbf{k}}^{\mu} = \frac{\pi}{2}\left|\widetilde{J}^{\text{odd}}\right|^{2} \sum_{\mathbf{q},\eta\nu} \left|\widetilde{\Phi}_{\mathbf{q},\mathbf{k}-\mathbf{q};\mathbf{k}}^{\eta\nu\mu}\right|^{2} \delta(\varepsilon_{\mu\mathbf{k}} - \varepsilon_{\eta\mathbf{q}} - \varepsilon_{\nu\mathbf{k}-\mathbf{q}}). \tag{17}$$

We take $\widetilde{J}^{\text{odd}}$ as an average value from (16), and $\widetilde{\Phi}_{\mathbf{q},\mathbf{k}-\mathbf{q};\mathbf{k}}^{\eta\nu\mu}$ is a dimensionless function of eigenvectors from the diagonalization of the quadratic Hamiltonian.

The decay rate is related to a much simpler quantity, the on-shell two-magnon density of states (DOS),

$$D_{\mathbf{k}}^{\mu} = D_{\mathbf{k}}(\varepsilon_{\mu\mathbf{k}}) = \pi \sum_{\mathbf{q},\nu\eta} \delta\left(\varepsilon_{\mu\mathbf{k}} - \varepsilon_{\nu\mathbf{q}} - \varepsilon_{\eta\mathbf{k}-\mathbf{q}}\right). \tag{18}$$

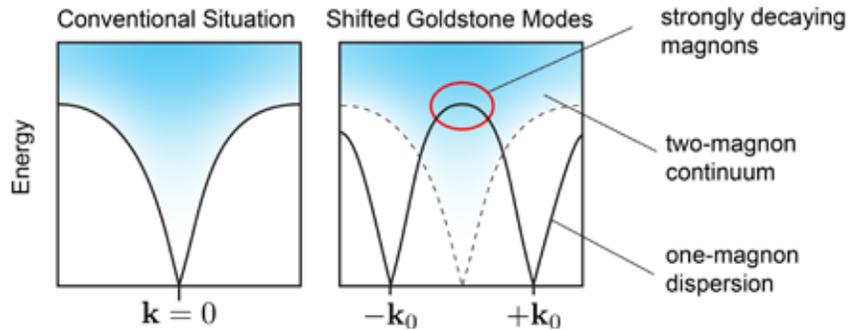

SUPPLEMENTARY FIGURE 3. **Kinematic conditions for two-magnon decay.** This sketch shows the kinematic conditions for two-magnon decay emphasizing the importance of the (pseudo-) Goldsone mode at a finite $\mathbf{Q}$-value.

9

We propose to approximate the square of the dimensionless vertex $\big|\widetilde{\Phi}^{\eta\nu\mu}_{\mathbf{q},\mathbf{k}-\mathbf{q};\mathbf{k}}\big|^2$ in the decay rate (17) by a constant, thus eliminating the numerically complex and costly element of the calculation and bypassing its analytical cumbersomeness. As a result, the decay rate (17) is simply proportional to the on-shell two-magnon DOS (18)

$$\gamma^{\mu}_{\mathbf{k}} \approx \frac{f}{2} \big|\widetilde{\mathcal{J}}^{\text{odd}}\big|^2 D^{\mu}_{\mathbf{k}}, \tag{19}$$

with $f = \langle \big|\widetilde{\Phi}^{\eta\nu\mu}_{\mathbf{q},\mathbf{k}-\mathbf{q};\mathbf{k}}\big|^2 \rangle$ and brackets implying averaging. We argue that this approximation is well-justified for the gapped systems and for models with lower spin symmetries. This is because Bogolyubov transformations are not singular without the true Goldstone modes and because the lower symmetry implies fewer restrictions on the decay amplitudes beyond the kinematic constraints in the DOS, see [26] and [30].

Another strong à posteriori justification of this procedure comes from the need of a regularization of the imprints of the two-magnon Van Hove singularities in the Born decay rate $\gamma^{\mu}_{\mathbf{k}}$ (17) that are inherited from the two-magnon DOS (18). Such singularities are unavoidable, see [26], yet they are unphysical and must be regularized. Regularization procedures result in an averaging of singularities over the momentum space, thus complementing the averaging suggested in our approximation (19). We also note that performing such a regularization in case of the fully numerical calculation of the vertex $\widetilde{\Phi}^{\eta\nu\mu}_{\mathbf{q},\mathbf{k}-\mathbf{q};\mathbf{k}}$ can be prohibitively costly.

The method that we use to address the issue of singularities is referred to as the iDE method, e.g. [31, 32]. It is a simplified version of Dyson's equation on the pole with only imaginary part of the equation solved self-consistently, which is in line with the approximations described above, $\Sigma_{\mu\mathbf{k}}(\varepsilon_{\mu\mathbf{k}} + i\gamma^{\mu}_{\mathbf{k}}) = -i\gamma^{\mu}_{\mathbf{k}}$. Allowing the initial magnon to have a finite lifetime relaxes the energy and momentum conservations, thus removing singularities of the Born approximation and mitigating the unphysical largeness of the $\gamma^{\mu}_{\mathbf{k}}$'s. Technically, instead of the integral in (17) with a simplifying assumption of (19), the calculation of $\gamma^{\mu}_{\mathbf{k}}$ now requires a recursive solution of

$$1 = \frac{f}{2} \sum_{\mathbf{q},\nu\eta} \frac{\big|\widetilde{\mathcal{J}}^{\text{odd}}\big|^2}{(\varepsilon_{\mu\mathbf{k}} - \varepsilon_{\nu\mathbf{q}} - \varepsilon_{\eta\mathbf{k}-\mathbf{q}})^2 + (\gamma^{\mu}_{\mathbf{k}})^2}, \tag{20}$$

which is, typically, a quickly convergent process. To provide a reasonable estimate of the constant $f$, we use our previous study of the XXZ model on the same (honeycomb) lattice in external field [30], for which analytical expressions for the dimensionless cubic vertices $\widetilde{\Phi}^{\eta\nu\mu}_{\mathbf{q},\mathbf{k}-\mathbf{q};\mathbf{k}}$ can be obtained. In this model [30], the characteristic values of the three-magnon couplings $\widetilde{\mathcal{J}}^{\text{odd}}$ relative to the magnon bandwidth are close to the ones considered in the current study, single-magnon states significantly overlap with the high-intensity parts of the two-magnon continuum in high fields, also in a close similarity to the kinematics of the present work, and magnon decays were shown to be very significant. By comparing Born-approximation $\gamma^{\mu}_{\mathbf{k}}$ for the XXZ model with the corresponding two-magnon density of states [30], we extract the value of $f \simeq 1/9$. Back-of-the-envelope estimates suggest this constant to be $f \simeq 4/zn^2$, where $z$ is the coordination number and $n$ is the number of magnon branches (sites in the unit cell). This estimate comes from analyzing the structure of the dimensionless cubic vertices $\widetilde{\Phi}^{\eta\nu\mu}_{\mathbf{q},\mathbf{k}-\mathbf{q};\mathbf{k}}$ in previous studies such as Supplementary Ref. [30]. As is clear from Supplementary Eq. (14), the real-space coupling of spin-flips affects nearest-neighbour Holstein-Primakoff magnons. Hence, the vertex in $\mathbf{k}$-space contains an

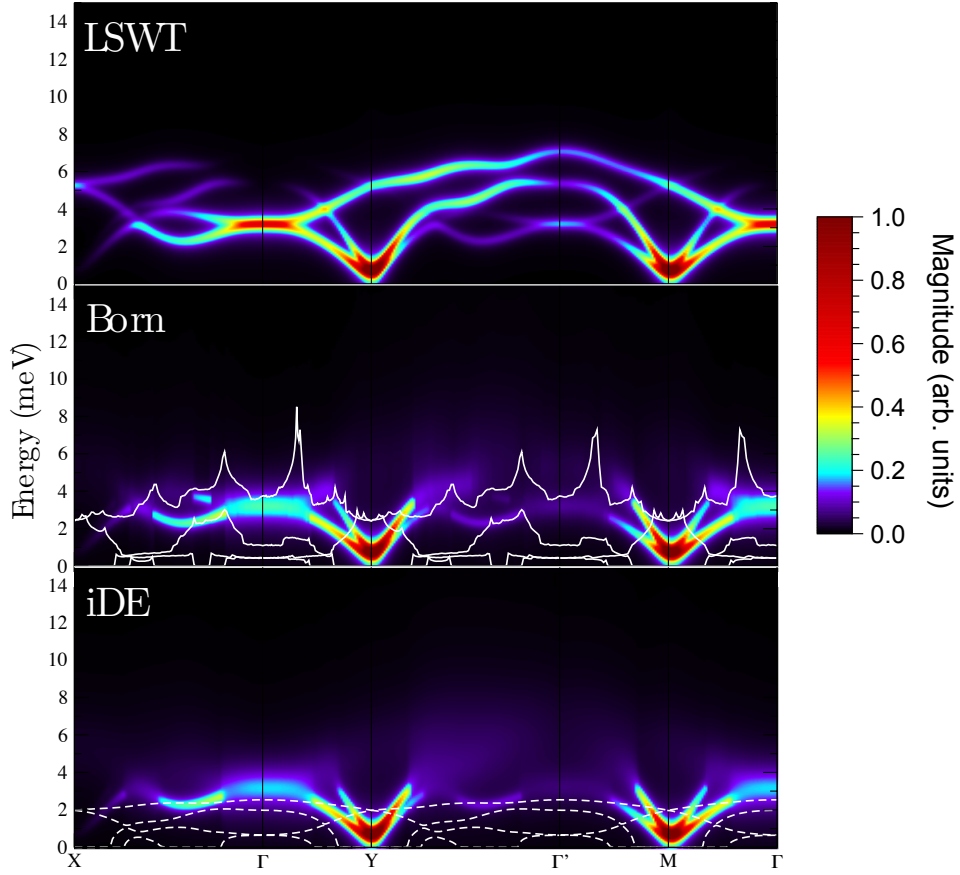

SUPPLEMENTARY FIGURE 4. **Comparison of LSWT, Born and iDE.** The ordering vector in the $\mathcal{I}(\mathbf{q},\omega)$ intensity plots was chosen to be at the M'-point $[(-\pi,\pi/\sqrt{3})]$. Upper panel: Linear spin wave theory (LSWT) along the contour XΓYΓ'MΓ in Fig. 3(b) of the main text with artificial broadening $\delta = 0.25$ meV. Middle panel: same with broadening $\gamma_{\mathbf{q}}^{\mu}$ from Supplementary Eq. (19) representing the approximate Born expression. Lower panel: same with broadening $\gamma_{\mathbf{q}}^{\mu}$ from Supplementary Eq. (20) representing the self-consistent imaginary Dyson equation (iDE) approach. Solid and dashed lines are $\gamma_{\mathbf{q}}^{\mu}$ from (19) and (20), respectively, for the four magnon modes with the higher values corresponding to the higher-energy modes.

analog of the nearest-neigbour hopping matrix. Averaging of its square yields with $\sim 1/z$ the inverse coordination number. The number of atoms in the magnetic unit cell $n$ gives the number of independent magnon modes and, therefore, their wavefunctions are normalized by $1/\sqrt{n}$. Since the vertex couples three magnons, its square is, thus, proportional to $\sim 1/n^3$. The summation over such modes eliminates one power of $n$. The factor of 4 comes from the square of the symmetrization factor in the decay term. For the considered problem of $n = 4$ and $z = 3$ the value of $f \simeq 1/12$ is in a quantitatively close agreement with the value of $f \simeq 1/9$.

**Dynamical Structure Factor**

Our results for the dynamical structure factor $\mathcal{I}(\mathbf{q},\omega)$ for Model 2 are presented in Supplementary Figs. 4 and 5. Supplementary Figure 4 shows $\mathcal{I}(\mathbf{q},\omega)$ for the zigzag state with

11

the ordering vector at the M′-point $[(-\pi, \pi/\sqrt{3})]$, with that choice motivated by a close similarity of $\mathcal{I}(\mathbf{q},\omega)$ along the XKΓYΓ′MΓ **k**-path (see Fig. 3(c) of the main text) with the one averaged over three zigzag configurations, shown in Supplementary Figure 5.

The upper panels in both Supplementary Figures show the LSWT results with an artificial Lorentzian broadening of $\delta = 0.25$ meV. The middle and the lower panels in Supplementary Figure 4 and the two lower panels in Supplementary Figure 5 show $\mathcal{I}(\mathbf{q},\omega)$ with the broadened magnon lines according to

$$\delta(\omega - \varepsilon_{\mu\mathbf{q}}) \to A_\mu(\mathbf{q},\omega) = \frac{1}{\pi}\frac{\gamma^\mu_\mathbf{q}}{(\omega - \varepsilon_{\mu\mathbf{q}})^2 + (\gamma^\mu_\mathbf{k})^2}, \qquad (21)$$

where the broadening is obtained from the approximate Born expression (19) for the middle panel of Supplementary Figure 4 and by the iDE method (20) described above for the rest of the panels. The second panel in Supplementary Figure 5 shows exact diagonalization results from Fig. 3(a) of the main text.

The solid lines in the middle panel of Supplementary Figure 4 show Born approximation decay rates $\gamma^\mu_\mathbf{q}$ from (18) with the three-magnon coupling $\widetilde{J}^{\text{odd}}$ from (16) and $f = 1/9$ as discussed above. For each **k**-point, there are four values of $\gamma^\mu_\mathbf{q}$, one for each branch, with the larger value corresponding to the magnon that is higher in energy. With $\gamma^\mu_\mathbf{q}$ reaching about a half of the total magnon bandwidth, there is hardly anything visible left from the intensity of the highest-energy mode. The second highest mode is also overdamped in most of the **k**-space. This is in agreement with the high-energy magnons having a significant two-magnon continuum phase space for decays.

The two lower-energy modes are also broadened, with the modulations of the broadening following the **k**-regions where magnons do or do not overlap with the two-magnon continuum, such as in the vicinity of the Y and M points in Supplementary Figs. 4 and 5 for the latter case.

Last but not the least are the Van Hove singularities in $\gamma^\mu_\mathbf{q}$ from the two-magnon DOS. Their imprints are also visible in $\mathcal{I}(\mathbf{q},\omega)$ as sharp boundaries between more or less bright regions of intensity, hence more or less well-defined magnons. At **k**'s close to such singularities, the decay rates become unphysically large, violate the perturbative nature of the expansion, and need to be regularized.

Physically, since the Van Hove singularities are affecting magnons that are already within the two-magnon continuum and are, thus, decaying, the most relevant method for such a regularization is the iDE approach described above, which allows to self-consistently account for the broadening of the initial-state magnons.

The iDE broadening $\gamma^\mu_\mathbf{q}$ from (20) is shown in the lower panel of Supplementary Figure 4 by the dashed lines. As one can see, the iDE method yields smooth, completely regular decay rates, with their overall values decreased for the upper- and somewhat increased for the lower-energy modes. We emphasize again that this result is more realistic than that of the Born approximation, because the divergent behaviour violates the perturbative nature of the $1/S$ expansion and is unphysical.

Supplementary Figure 5 shows $\mathcal{I}(\mathbf{q},\omega)$ averaged over three zigzag configurations. Our iDE results in the first lower panel of Supplementary Figure 5 clearly capture many of the most notable features seen in the ED data and constitute a clear improvement over the LSWT results. There are still some notable differences. First, in the vicinities of the M- and Y-points, ED branches at lower energies are flatter and asymmetric with a more drastic decrease of intensity in the Y→ Γ′ direction. There is also only one mode resolved near

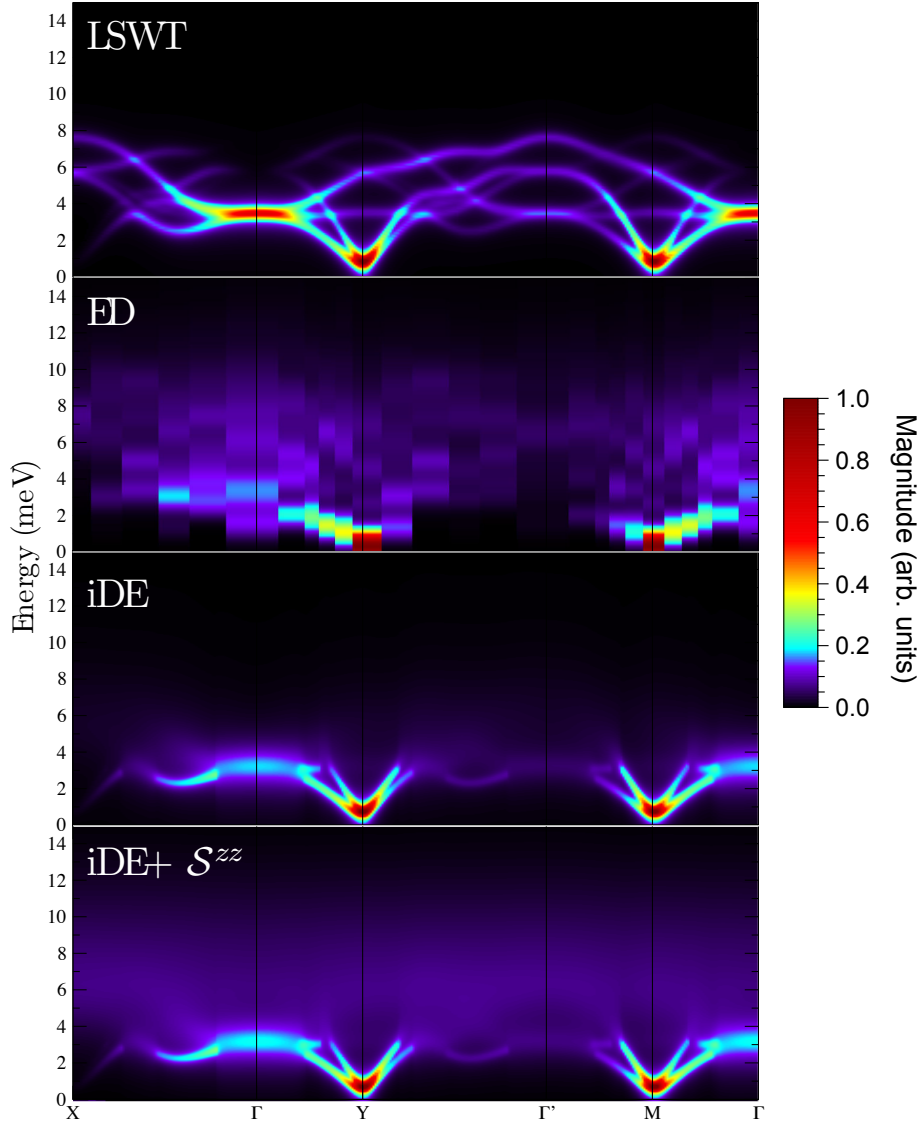

SUPPLEMENTARY FIGURE 5. **Averaged $\mathcal{I}(\mathbf{q},\omega)$ in various approaches.** $\mathcal{I}(\mathbf{q},\omega)$ intensity plots are averaged over the three zigzag states. Upper panel: LSWT with artificial broadening $\delta = 0.25$ meV. Second upper panel: ED results from Fig. 3(c) of the main text. First lower panel: LSWT with the iDE broadening $\gamma_{\mathbf{q}}^{\mu}$ from Supplementary Eq. (20). Second lower panel: same with the averaged longitudinal intensity $\mathcal{I}^{zz}(\mathbf{q},\omega)$, see text.

these points in the ED data, while the SWT predicts two. These may be ascribed to the ignoring of the real part of the self-energies, which neglects the spectrum flattening and the reduction of the quasiparticle peak intensity.

Another major remaining difference is the presence of a significant intensity in the ED data at the energies near and above the single-magnon band maximum, $\omega \gtrsim 7$ meV, which is completely missing in both LSWT and iDE results. This missing feature is beyond standard calculations of the structure factor, which take into account only transverse spin fluctuations (see Supplementary Ref. [33] for an exception). The missing part is the longitudinal component, $\mathcal{I}^{zz}(\mathbf{q},\omega)$ in the local $z$-axes, which is directly related to the continuum of the

13

broadened two-magnon excitations. While the full calculation of $\mathcal{I}^{zz}(\mathbf{q},\omega)$ is beyond the scope of the current work, a simple account of its $\omega$ structure is possible.

A naïve approach is to suggest a direct proportionality of $\mathcal{I}^{zz}(\mathbf{q},\omega)$ to the two-magnon DOS, similarly to Supplementary Eq. (19). However, a better approximation is achieved by modifying the density of states by including the broadenings of the magnon lines inside the continuum

$$\mathcal{I}^{zz}(\mathbf{q},\omega) = \frac{f_2}{2} \sum_{\mathbf{k},\mu\nu} \frac{\gamma_{\mathbf{k}}^{\mu} + \gamma_{\mathbf{k}-\mathbf{q}}^{\nu}}{(\omega - \varepsilon_{\mu\mathbf{k}} - \varepsilon_{\nu\mathbf{k}-\mathbf{q}})^2 + (\gamma_{\mathbf{k}}^{\mu} + \gamma_{\mathbf{k}-\mathbf{q}}^{\nu})^2}. \tag{22}$$

This modification is very physical and self-consistent as the continuum is built from the broadened magnons. It also eliminates sharp features in the continuum and because the $\gamma_{\mathbf{q}}^{\mu}$'s are larger for the upper magnon branches, the upper part of the continuum also gets washed out more. The single adjustable parameter $f_2$ can be argued to be $1/8$ using naïve estimates. While the resulting intensity lacks a more involved modulation in the momentum, one can expect an overall better description of the ED results.

The results shown in the lowest panel of Supplementary Figure 5 include $\mathcal{I}^{zz}(\mathbf{q},\omega)$ from (22) with $f_2 = 1/8$ and the averaged iDE decay rates: $\gamma_{\mathbf{q}}^1 = 0.5$ meV, $\gamma_{\mathbf{q}}^2 = 1.0$ meV, $\gamma_{\mathbf{q}}^3 = 1.5$ meV, and $\gamma_{\mathbf{q}}^4 = 2.5$ meV, where the numeration is from the lowest to the highest in energy.

Thus, the broad features in the full $\mathcal{I}(\mathbf{q},\omega)$ are a combination of the remnants of the broadened magnon modes from the transverse part of $\mathcal{I}(\mathbf{q},\omega)$ with the longitudinal $\mathcal{I}^{zz}(\mathbf{q},\omega)$ part, so the combination has a maximum at the energies between 6 and 7 meV, in a close resemblance of the ED data. While this is only an approximate description, it provides confidence that a complete account should be able to reproduce other features of the ED data in that range of energies as well.

Altogether, we believe we have been able to provide a convincing description of the most significant effects of the three-magnon interaction on the magnon spectrum that agrees with the numerical studies.

## SUPPLEMENTARY NOTE 4:
## FURTHER DETAILS OF EXACT DIAGONALIZATION CALCULATIONS

The exact diagonalization calculations in this work were carried out on the series of 20- and 24-site clusters with periodic boundary conditions shown in Supplementary Figure 6. As noted in the Methods section of the main text, ED calculations were performed using the Lanczos algorithm [34], employing the continued fraction method [35] to obtain the desired dynamical correlation functions. While the periodic cluster **24A** retains all the symmetries of Supplementary Eq. (1), the remaining clusters are of lower symmetry, resulting in slight anomalies in the symmetries of the computed correlation functions. For **24B**, **20A**, and **20B** the symmetry was partially restored by averaging the results over all symmetry-related orientations of the clusters, which generates the **k**-points shown in Supplementary Figure 6(b).

In the following pages, we show complete results obtained for various models. For each model, comparison of results for each cluster is shown for the high-symmetry $\Gamma$, X, M(Y) and $\Gamma'$ points, which live on all four clusters. These results appear in Supplementary Figure 8-10(d,h,l,p). For each case, the lowest energy peak positions are relatively well converged with respect to finite-size effects (compared to the chosen 0.5 meV Gaussian broadening). For

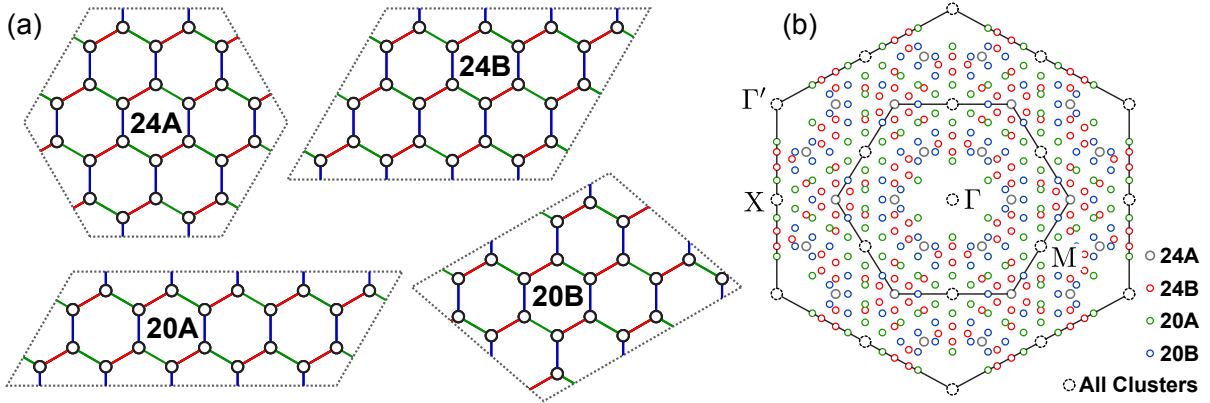

SUPPLEMENTARY FIGURE 6. **Cluster details in real and reciprocal space.** (a) 20- and 24- site periodic clusters employed for exact diagonalization (ED) studies in this work. (b) Summary of **k**-points associated with each cluster. For the low-symmetry **20A**, **20B**, and **24B**, **k**-points are shown for all symmetry-related orientations of the cluster. The high-symmetry points $\Gamma$, X, M(Y) and $\Gamma'$ live on all clusters.

excitations representing a continuum, we observe variations in the positions of the higher energy peaks obtained from the various clusters, as might be expected. Averaging over the discrete excitations of the different clusters therefore restores the continuum, improving the validity of the computed intensities at the high-symmetry $\Gamma$, X, M(Y) and $\Gamma'$ points. However, we note that away from the high-symmetry points, where averaging is not possible, the intensities are less reliable. This observation does not alter the conclusions drawn from plotting $\mathcal{I}(\mathbf{k}, \omega)$ along various **k**-paths (as in Supplementary Figure 7), but should be noted. In particular, in order to avoid spurious features, the plots of the **k**-dependence of $\mathcal{I}(\mathbf{k}, \omega)$ for various energy intervals (Fig. 2(d) and 3(d), main text, and Supplementary Figure 8-10(a,e,i,m)) employed only data from the highest symmetry **24A** cluster. Plots of $\mathcal{I}(\mathbf{k}, \omega)$ along the particular **k**-path (Fig. 2(c) and 3(c), main text, and Supplementary Figure 8-10(b,f,j,n)) employed data from all clusters. In Supplementary Figure 7 we identify the periodic cluster associated with each **k**-point.

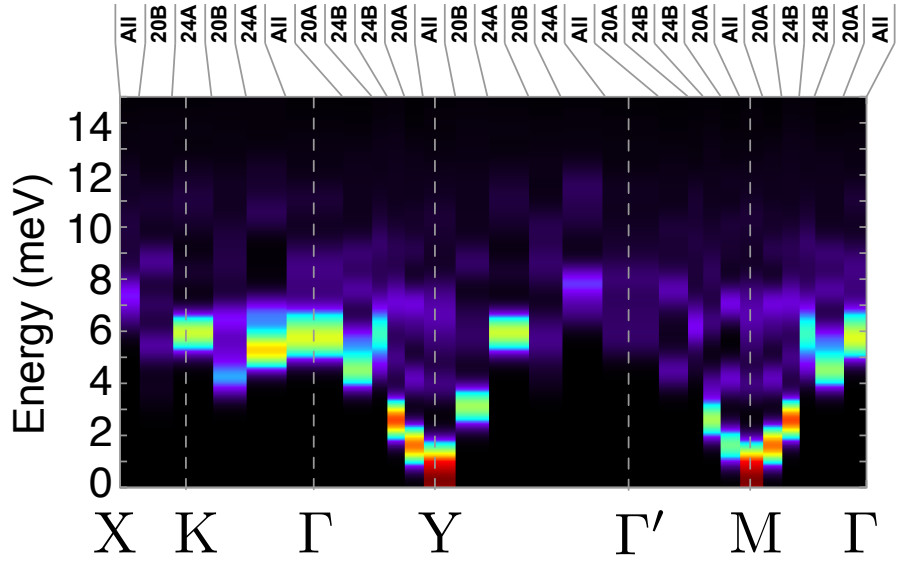

SUPPLEMENTARY FIGURE 7. **Specific contributions of periodic clusters.** Predicted $\mathcal{I}(\mathbf{k},\omega)$ for $J_1 = -4.6, K_1 = +7.0$ meV showing the periodic cluster of origin for each momentum interval. For intervals labelled "All" the obtained intensities were averaged over the four clusters. Comparison of the intensities at such **k**-points is shown in Supplementary Figure 8-10(d,h,l,p).

# SUPPLEMENTARY NOTE 5:
# ADDITIONAL ED RESULTS FOR VARIOUS MODELS

In this section, we show full results for various additional parameters not appearing in the main text, as well as a comparison of results from the different periodic clusters **20A-24B**.

**Nearest Neighbour Heisenberg-Kitaev (nnHK) Model**

We first show, in Supplementary Figure 8, results obtained for the nnHK model with $K_1 > 0$ and $J_1 < 0$. We begin with $J_1 = -4.6$ meV, $K_1 = +7.0$ meV (Supplementary Figure 8(a-d)), as suggested from analysis of powder inelastic neutron scattering data in Supplementary Ref. [22], and then consider several models moving towards the spin-liquid, maintaining constant $\sqrt{K_1^2 + J_1^2}$.

The validity of the ED approach can be seen by comparing the results for the pure Kitaev model (Supplementary Figure 8(n)) with the exact results (Supplementary Figure 8(o)). One can see good agreement between $\mathcal{I}(\mathbf{k}, \omega)$ predicted from the two approaches. A similar degree of agreement is seen for $J_1 = -4.6$ meV, $K_1 = +7.0$ meV, for which the ED results (Supplementary Figure 8(b)) and the LSWT results (Supplementary Figure 8(c)) are also in close correspondence. This model exists close to the hidden SU(2) point (at $K_1 = -2J_1$) noted in Supplementary Ref. [7]. For this reason, the dynamics are expected to be well described by conventional spin-waves. Therefore, the LSWT method performs well both close to and far away from the spin liquid.

As discussed in the main text, upon approaching the spin-liquid, the intensive excitations at the 2D Γ-point shift to higher energy, and become increasingly broad as they move deeper into the three-magnon continuum. Away from the Γ-point, the intensive excitations are mostly associated with the lowest magnon band at the level of LSWT. These excitations remain relatively sharp in the ED calculations, and shift to lower energy on approaching the spin-liquid. This effect is clearly seen for the computed intensity at the X-point, shown in Supplementary Figure 8(d,h,l,p), which remains sharply peaked over the entire studied range. As discussed in the main text and Supplementary Note 3, low-energy magnons in the nnHK model are protected from decay due to the absence of three-magnon states at low energies. As shown in Supplementary Figure 8(n,o), these low energy excitations eventually evolve into a flat band in the spin liquid with intensity peaked just above the two-flux gap $\Delta \sim 0.065|K_1|$ [12, 13].

As in Supplementary Ref. [36], we find that the best agreement with the experimental results within the nnHK model is obtained for $|J_1/K_1| = 0.3$, which shows a star-like pattern at intermediate energy. This ratio of interactions was featured in the main text, for Model 1. However, we also find significant intensity at the X-points, inconsistent with the observed intensities. Moreover, as noted in the main text, the absence of low-energy intensity at the Γ-point in these models is strongly inconsistent with the experimental data of Supplementary Refs. [17, 19].

SUPPLEMENTARY FIGURE 8. **Detailed results for the nnHK model.** For each model we show, for example, (a) **k**-dependence of intensity integrated over the indicated energy regions, (b) ED and (c) LSWT intensities over the indicated **k**-path. (d) Comparison of the intensities obtained from the various clusters; intensities are not shown to scale between different **k**-points. Analogous results are shown in (e)-(p). For the spin-liquid, (m) and (o) represent exact results (see Supplementary Refs. [12, 13]).

18

**Extended $(J_1, K_1, \Gamma_1, J_3)$ Model, $K_1 > 0$**

We next consider parameters in the region suggested by *ab initio* studies of the earlier $P3_112$ structure of $\alpha$-RuCl$_3$: $|J_1| \sim |K_1| \sim |\Gamma_1|$, with $J_1 < 0, K_1 > 0, \Gamma_1 > 0$ [8, 21]. It is worth noting that this region of the phase diagram features competition between ferromagnetic, zigzag, and 120° order, such that the combination of $K_1 > 0$ and a finite $\Gamma_1 > 0$ tends to destabilize the zigzag order. In order to restore the zigzag ground state, we therefore add a small $J_3$ coupling. We start with the interactions suggested in Supplementary Ref. [22], namely $J_1 = -4.6$ meV, $K_1 = +7.0$ meV. We then add a small $J_3 = 0.7$ meV, and then vary the ratio of $|\Gamma_1/K_1|$, holding constant $J_1, J_3$, and $\sqrt{J_1^2 + K_1^2 + \Gamma_1^2 + J_3^2}$. Results are shown in Supplementary Figure 9.

Comparing Supplementary Figure 9(a-d) with Supplementary Figure 9(e-h), one can see that the addition of a small $J_3$ does not significantly influence the spectra. However, similar to the $K_1 < 0$ region, we observe significant broadening of the ED spectra upon increasing $\Gamma_1$. For $(J_1, K_1, \Gamma_1, J_3) = (-4.6, +5.0, +5.0, +0.7)$ meV (Supplementary Figure 9(m-p)), LSWT predicts a large spin-wave gap, and relatively flat dispersion for the spin-wave bands. Interestingly, the momentum dependence of the predicted intensities from LSWT resemble somewhat those of the $K_1 < 0$ Kitaev spin-liquid, but shifted to higher energy. That is, there appears a flat band at 4-5 meV, with intensity centered around the 2D $\Gamma$-point, and another band at higher energies 10-12 meV, with intensity away from the $\Gamma$-point. The vanishing dispersion of these bands at the level of LSWT is likely related to close proximity to the phase boundaries between ferromagnetic, zigzag, and 120° order, which would typically feature low-energy modes near the $\Gamma$, M(Y), and K-points, respectively.

Results for the ED calculations differ significantly from the LSWT intensities at large $\Gamma_1$. In particular, the gap is significantly reduced, such that dispersing modes can be observed near the (M,Y)-points. This may result from shifting of the phase boundaries in the ED calculations compared to the semiclassical LSWT approach. Interestingly, for $(J_1, K_1, \Gamma_1, J_3) = (-4.6, +5.0, +5.0, +0.7)$ meV (Supplementary Figure 9(m-p)) we observe low-energy intensity at the $\Gamma$-point in the ED calculations, and a star-like shape at intermediate energies, consistent with the experimental data on $\alpha$-RuCl$_3$. However, in the high energy region, the intensity is mainly located away from the $\Gamma$-point, in contradiction with the experiment of Supplementary Ref. [19].

SUPPLEMENTARY FIGURE 9. **Detailed results for the extended model with $K_1 > 0$.** For each model we show, for example, (a) **k**-dependence of intensity integrated over the indicated energy regions, (b) ED and (c) LSWT intensities over the indicated **k**-path. (d) Comparison of the intensities obtained from the various clusters; intensities are not shown to scale between different **k**-points. Analogous results are shown in (e)-(p).

**Extended $(J_1, K_1, \Gamma_1, J_3)$ Model, $K_1 < 0$**

Finally, we show, in Supplementary Figure 10, complete results for a variety of models in the region suggested by various *ab initio* studies based on recent $C2/m$ structures of $\alpha$-RuCl$_3$, that is $J_1 \sim 0, K_1 < 0, \Gamma_1 > 0, J_3 > 0$ [8, 10, 23–25]. In each case, we hold $J_1 = -0.5$ meV, $J_3 = +0.5$ meV and the overall scale $\sqrt{J_1^2 + K_1^2 + \Gamma_1^2 + J_3^2}$ constant, and modify the ratio of $|K_1/\Gamma_1|$. We also show results for the $K_1 < 0$ Kitaev spin-liquid for comparison.

As discussed in the main text and Supplementary Note 3, a large $\Gamma_1$ interaction induces significant deviations between the ED and LSWT results, due to coupling between the one- and two-magnon excitations. For Supplementary Figure 10(b,f), only the excitations around the (M,Y)-points are sharply peaked in the ED calculations. On approaching the spin-liquid (by decreasing $|\Gamma_1/K_1|$), the continuum is shifted to lower energies, and becomes sharper, evolving into the flat sharp band in the spin-liquid. The importance of $\Gamma_1$ interactions can further be seen by comparing the results of Supplementary Figure 8(a-d) with Supplementary Figure 10(a-d). For these models, the LSWT predictions for $\mathcal{I}(\mathbf{k}, \omega)$ are very similar despite remarkably different interaction parameters. However, the ED results differ substantially.

As noted in the main text and Supplementary Note 3, the requirements for strong coupling of the one- and two-magnon excitations include a deviation of the ordered moments from the high symmetry cubic axes. While finite $\Gamma_1$ interactions generally rotate the ordered moments away from the cubic axes, it is interesting to consider also the case where $\Gamma_1 = 0$, $J_3 > 0$, and $K_1 < 0$. In this case, the directions of the ordered moments are not completely determined at the classical level. For example, for the ordering wavevector $\mathbf{Q}$ parallel to the $Z$ bond, the classical energy is minimized for any orientation of the moments in the $xy$-plane. However, as noted in Supplementary Refs. [27, 37] the cubic axes are selected by a quantum order-by-disorder mechanism, such that magnons are expected to remain stable in this limit. Indeed, comparison of the ED and LSWT results in Supplementary Figure 10(j) with (k) shows general agreement. This observation further establishes the importance of off-diagonal anisotropic interactions such as $\Gamma_1$.

Finally, we note that the models studied in this region (such as Model 2 of the main text) display excitation gaps on the order of 0.5 meV (at the M(Y)-points) at the level of both ED and LSWT. This is in contrast with the neutron scattering results, which appear to show a gap on the order of 2 meV [22, 38]. However, it is worth noting that the size of the excitation gap is strongly influenced by the relative magnitudes of the $K_1$ and $\Gamma_1$ interactions along each nearest neighbour $X,Y$ and $Z$ bond, which are not constrained to be equal by the symmetry of the real crystals [8, 10]. To demonstrate this, we show LSWT results for Model 2 (Supplementary Figure 11), modified with an anisotropic $K_1$ and $\Gamma_1$, consistent with the results of Supplementary Ref. [8]. Specifically, we show $J_1 = -0.5$, $J_3 = +0.5$, with $K_1^Z = -5.0 + \delta, K_1^{XY} = -5.0 - \delta, \Gamma_1^Z = +2.5 + \delta/2, \Gamma^{XY} = +2.5 - \delta/2$. The gap can be reproduced already for small perturbations on the order of $\delta = 0.1 K_1$, while the remainder of the dispersions are not strongly affected. Furthermore, additional small anisotropic interactions [39–41] are allowed by symmetry that may also contribute to the gap. For simplicity, we have neglected such additional terms in the main text, but expect that their inclusion would further improve agreement with the experimental observations.

SUPPLEMENTARY FIGURE 10. **Detailed results for the extended model with $K_1 < 0$.** For each model we show, for example, (a) **k**-dependence of intensity integrated over the indicated energy regions, (b) ED and (c) LSWT intensities over the indicated **k**-path. (d) Comparison of the intensities obtained from the various clusters; intensities are not shown to scale between different **k**-points. Analogous results are shown in (e)-(p). For the spin-liquid, (m) and (o) represent exact results (see Supplementary Refs. [12, 13]).

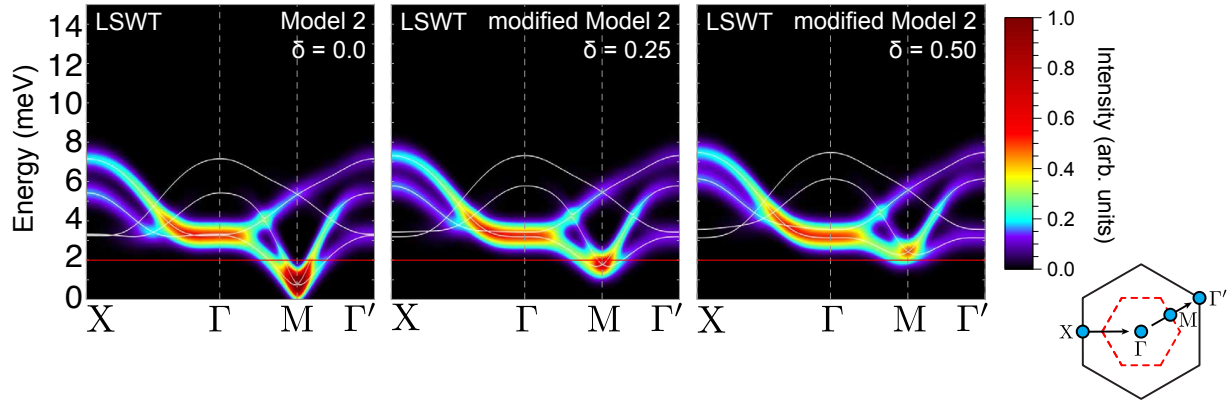

SUPPLEMENTARY FIGURE 11. **Bond-dependent interactions: Evolution of the gap.** Bond-dependent interactions are introduced at the LSWT level to investigate the evolution of the gap at the M-point. The ordering wavevector is the Y-point in each case, and results are not averaged over 120 degree domains. The experimental value of $\sim 2.0$ meV is indicated by a red line. Relatively small perturbations are sufficient to reproduce the experimental gap.


\end{document}